\documentclass[aps,prd,nofootinbib,showkeys,showpacs,superscriptaddress,preprint,onecolumn,11pt]{revtex4-2}

\usepackage{graphicx}
\usepackage{dcolumn}
\usepackage{bm}
\usepackage{amsmath,amsfonts,amsthm,amssymb}
\usepackage{hyperref} 
\usepackage{float}
\usepackage{lipsum}
\usepackage{physics}
\usepackage{nicefrac}
\usepackage{yfonts}
\usepackage[utf8]{inputenc}
\usepackage{cancel}

\usepackage[dvipsnames]{xcolor}

\newcommand{\Ckv}{{\boldsymbol C}_k} 
\newcommand{\Cov}{{\boldsymbol C}_\omega} 

\newcommand{\mgoth}{\mathfrak{m}}
\newcommand{\pgoth}{\mathfrak{p}}

\usepackage[sort&compress]{natbib}
\hypersetup{
colorlinks = true,
linkcolor = blue,
anchorcolor = blue,
citecolor = blue,
filecolor = blue,
urlcolor = blue
}
\usepackage[normalem]{ulem}
\usepackage{orcidlink}

\newcommand{\bea}{\begin{eqnarray}}
\newcommand{\eea}{\end{eqnarray}}
\newcommand{\bel}[1]{\begin{eqnarray}\label{#1}}
\newcommand{\eel}{\end{eqnarray}}

\def\LB{\left(}
\def\RB{\right)}
\def\LSB{\left[}
\def\RSB{\right]}

\newcommand{\nn}{\nonumber}

\newcommand{\EQ}[1]{Eq.~(\ref{#1})}
\newcommand{\EQn}[1]{(\ref{#1})}

\newcommand{\EQSTWO}[2]{Eqs.~(\ref{#1})~and~(\ref{#2})}

\newcommand{\EQSM}[2]{Eqs.~(\ref{#1})--(\ref{#2})}

\newcommand{\FIG}[1]{Fig.~\ref{#1}}

\newcommand{\CIT}[1]{Ref.~\citep{#1}} 
\newcommand{\CITn}[1]{\citep{#1}}

\newcommand{\spin}{\mathfrak{s}}

\newcommand{\Uvv}{{\boldsymbol U}}

\newcommand{\uv}{{\boldsymbol u}}
\newcommand{\vv}{{\boldsymbol v}}
\newcommand{\pvv}{{\boldsymbol p}}

\def\be{\begin{equation}}
	\def\ee{\end{equation}}
\def\barr{\begin{array}}
	\def\earr{\end{array}}
\def\beq{\begin{eqnarray}}
	\def\eeq{\end{eqnarray}}
\def\bfig{\begin{figure}}
	\def\efig{\end{figure}}

\newcommand{\p}{\partial}

\newcommand{\f}[2]{\frac{#1}{#2}}

\def\g5{\gamma_5}

\def\S0iU{{\Sigma}^{0i}}

\def\n0{n_{(0)}}
\def\e0{\varepsilon_{(0)}}
\def\P0{P_{(0)}}

\begin{document}

\title{Boost-invariant and cylindrically symmetric perfect spin hydrodynamics}
            
\author{Zbigniew Drogosz\orcidlink{0000-0001-5133-958X}}
\email{zbigniew.drogosz@alumni.uj.edu.pl}
\affiliation{Institute of Theoretical Physics, Jagiellonian University, Kraków, 30-348, Poland}  
          
\author{Wojciech Florkowski\orcidlink{0000-0002-9215-0238}}
\email{wojciech.florkowski@uj.edu.pl}
\affiliation{Institute of Theoretical Physics, Jagiellonian University, Kraków, 30-348, Poland}   

\author{Jakub Witkowski\orcidlink{0009-0007-5310-8356}}
\email{jakub.witkowski@student.uj.edu.pl}
\affiliation{Institute of Theoretical Physics, Jagiellonian University, Kraków, 30-348, Poland} 

\begin{abstract}
Equations of a boost-invariant and cylindrically symmetric perfect hydrodynamics are solved numerically for initial conditions inspired by the wounded nucleon model. The energy--momentum and spin tensors are used in the form that describes a relativistic massive gas governed by Boltzmann statistics. In contrast to one dimensional boost-invariant expansion, we find a coupling between the azimuthal and longitudinal components of the electric and magnetic components of the spin polarization tensor. This feature is similar to that found earlier for the Gubser symmetry, however, our treatment allows for a more general form of initial conditions and expansion geometry. Defining the freeze-out hypersurface by the constant temperature condition, we evaluate the Pauli--Lubański four-vector and find that for the assumed geometry the only nonzero total polarization may be induced by the longitudinal component of the magnetic part of the spin polarization tensor coupled with the azimuthal electric component. The obtained results may serve as a reference point for more realistic models of hydrodynamic expansion.
\end{abstract}

\maketitle

\section{Introduction}
\label{introduction}

The experimental measurements of hyperon spin polarization~\cite{STAR:2017ckg, STAR:2018gyt, STAR:2019erd,ALICE:2019onw,STAR:2020xbm,ALICE:2021pzu,STAR:2021beb,HADES:2022enx,STAR:2023nvo} and vector-meson spin alignment~\cite{ALICE:2019aid,STAR:2022fan,ALICE:2022dyy} inspired a great deal of interest in the spin phenomena that may occur in the collisions of heavy ions at relativistic energies. For recent reviews, see~\cite{Becattini:2024uha, Niida:2024ntm, Chen:2024afy}, and for various related theoretical studies, see~\cite{Liang:2004ph, Liang:2004xn, Becattini:2009wh, Becattini:2013fla, Yang:2017sdk, Montenegro:2017rbu, Montenegro:2017lvf, Florkowski:2017ruc, Florkowski:2017dyn, Becattini:2017gcx, Florkowski:2018ahw, Sun:2018bjl, Sheng:2019kmk, Hattori:2019lfp, Xie:2019jun, Weickgenannt:2019dks, Florkowski:2019voj, Weickgenannt:2020aaf, Bhadury:2020cop,Shi:2020htn, Yi:2021ryh, Hongo:2021ona, Wang:2021wqq, Liu:2021uhn, Fu:2021pok, Becattini:2021suc, Becattini:2021iol, Sheng:2022wsy,Sheng:2022ffb,Hongo:2022izs,Weickgenannt:2022zxs, Sarwar:2022yzs, Wagner:2022amr,Wagner:2022gza, Kumar:2022ylt, Biswas:2023qsw, Weickgenannt:2023btk, Kiamari:2023fbe,Fang:2023bbw,Palermo:2023cup, Palermo:2023qfg,Kumar:2023ojl, Kumar:2023ghs, Chen:2023hnb,Goncalves:2024xzo,Grossi:2024pyh,Sun:2024anu, Singh:2024cub, Banerjee:2024xnd, Sapna:2025yss, Grossi:2025hxg, Arslan:2025tan, Liu:2025klr, Florkowski:2026ofs}.

One of the important theoretical developments related to this topic is the construction of spin hydrodynamics. In this paper, we analyze some formal aspects of this theory, namely, we construct its boost invariant and cylindrically symmetric  solutions. We restrict ourselves to the case of perfect spin hydrodynamics in the form defined in our recent papers~\CITn{Bhadury:2025dzh}, i.e., we assume that the spin part of the total angular momentum is separately conserved and adopt the so-called GLW (de Groot--van Leeuwen--van Weert) versions of the energy--momentum and spin tensors~\CITn{DeGroot:1980dk, Florkowski:2018fap}. Moreover, we do not include the feedback of the spin variables on the hydrodynamic evolution of the spinless background. 

Our current analysis can be viewed as an extension of earlier works that included a one-dimensional boost invariant expansion \cite{Florkowski:2019qdp, Florkowski:2021wvk,  Singh:2021man, Wang:2021ngp, Biswas:2022bht, Drogosz:2024lkx, Drogosz:2026qbo} and a Gubser hydrodynamic flow \CITn{Singh:2020rht, Singh:2026wvf}; see also~\CITn{Rindori:2021quq, Weickgenannt:2024esg, Singh:2026ytd, Palermo:2026mwu}. In contrast to \CITn{Singh:2020rht, Singh:2026wvf, Li:2026vld}, our transverse dynamics is not constrained by any symmetry except the cylindrical one. The solutions in the transverse plane are obtained numerically for physically motivated initial conditions -- since cylindrical symmetry is appropriate for central collisions, the initial transverse profiles of the entropy density are expressed by the Woods--Saxon nucleon distributions.

With our two symmetries constraining the system dynamics, we cannot address the two phenomena which are commonly discussed in the context of spin physics, namely, the global and local longitudinal polarizations. The global polarization is connected with the rotation about the $y$-axis (out-of-plane axis) which is absent in boost invariant systems, whereas the longitudinal polarization is connected with the formation of the elliptic flow and does not appear in a cylindrically symmetric case. However, in our opinion, our analysis is important for the very development of spin hydrodynamics, since very little is known about the properties of its solutions. On the one hand side, one-dimensional boost invariant and two dimensional Gubser solutions have been constructed, and on the other side the first numerical solution have been found for a full 3D dynamics~\CITn{Singh:2024cub, Sapna:2025yss}. Our present study is the first attempt to fill this gap and analyze 2D dynamics, however, still with a constraint of cylindrical symmetry. Future, less restrictive analyzes may benefit from our findings, since they should be reduced to them in some special cases. 

The paper is organized as follows. In Sec.~\ref{sec:perfect} we define the structure of perfect spin hydrodynamics used in our analysis. In particular, we discuss splitting of the system dynamics into a spinless hydrodynamic background and a spin evolution in a given background.  Boost invariance and cylindrical symmetry are implemented in Sec.~\ref{sec:symmetries}. Section \ref{sec:hyd-back} describes the hydrodynamics equations of a spinless background. Numerical results are shown for the equation of state that describes a gas of massive particles. Equations describing the time evolution of the components of the spin polarization tensor are introduced in Sec.~\ref{sec:spinden}. In Sec.~\ref{sec:PLvector} we present our calculation of the Pauli--Lubański four-vector of particles with a given momentum, emitted from a system on the freeze-out hypersurface defined by the condition of constant temperature.  We summarize in Sec.~\ref{sec:summary}. Details of the calculations are given in the appendices.

\smallskip
{\it Notation and conventions}: For the Levi--Civita tensor
$\epsilon^{\mu\nu\alpha\beta}$ we follow the convention $\epsilon^{0123} =-\epsilon_{0123} = +1$. The metric tensor is of the form $g_{\mu\nu} = \textrm{diag}(+1,-1,-1,-1)$. We use boldface symbols to denote three vectors and their scalar products are written as $\bm{a}\cdot\bm{b}$. The particles are always considered on the mass shell with the energy $E_{\bm{p}} = \sqrt{m^2+\bm{p}^2}$. Throughout the text, we make use of natural units, $\hbar = c = k_{\rm B} = 1$.

\section{Perfect spin hydrodynamics}
\label{sec:perfect}

The perfect spin hydrodynamics is based on the conservation laws for the baryon number, energy, linear momentum, and the spin part of the total angular momentum (for a recent summary of this approach, see~\CITn{Bhadury:2025dzh}). These conservation laws take the differential form
\bel{eq:psh}
\p_\mu N^\mu  = 0, \quad \p_\mu T^{\mu \nu} = 0, \quad \p_\mu S^{\mu,\alpha \beta} = 0,
\eel
where $N^\mu$ is the baryon current, $T^{\mu \nu}$ is the energy--momentum tensor and $S^{\mu,\alpha \beta}$ is the spin tensor. Having in mind application to a gas of spin-1/2 particles in local equilibrium, we use the so called GLW versions of $T^{\mu \nu}$ and $S^{\mu,\alpha \beta}$~\CITn{Florkowski:2018fap}. In this case \EQn{eq:psh} becomes eleven equations for temperature ($T$), baryon chemical potential ($\mu$), three independent components of the fluid four-velocity ($\Uvv$), and six components of the spin polarization tensor ($\omega_{\mu\nu}$).

The GLW form of the energy--momentum tensor is symmetric, which implies that the corresponding GLW spin tensor is conserved. As a consequence, there is no transfer between the spin and orbital parts of the total angular momentum. Such a transfer is possible if we go beyond the description of a perfect fluid. In this case, the energy--momentum tensor receives asymmetric gradient corrections that imply that the divergence of the spin tensor does not vanish any longer. 

It is important to distinguish between gradient corrections that are dissipative and corrections that result from the expansion of various tensors in the magnitude of the spin polarization tensor $\omega$. The latter do not introduce dissipation in a way similar to that of expansion in the dimensionless ratio $\mu/T$ (low baryon-number density expansion). Since the expansion in $\omega$ for the baryon current and energy--momentum tensor starts with quadratic terms and starts with linear terms for the spin tensor, it is natural to consider the case where one first considers spinless hydrodynamic evolution and then solves the spin dynamics in a given hydrodynamic background. This corresponds formally to the expansion including only first order terms in $\omega$. This scheme is adopted in this work. An extension to higher orders in $\omega$ is planned for future works. 

\section{Symmetry implementation}
\label{sec:symmetries}

In this section, we implement cylindrical symmetry and boost invariance in the formalism of perfect spin hydrodynamics. We follow the treatment of~\CITn{Florkowski:2011jg, Florkowski:2019qdp} -- we first introduce cylindrically symmetric and boost-invariant basis four-vectors, and subsequently use them to construct the energy--momentum and spin tensors. 

\subsection{Choice of basis}
\label{sec:basis}

The hydrodynamic flow is described by the four-vector  
\begin{equation}
U^\mu = \gamma (1, \vv) = \gamma (1, v_x, v_y, v_z), \quad \gamma = (1-v^2)^{-1/2}.
\label{Umu}
\end{equation}
For boost-invariant and cylindrically symmetric systems, we may use the following parameterization~\CITn{Florkowski:2011jg, Florkowski:2019qdp}
\begin{eqnarray}
U^0 &=& \cosh \theta \cosh \eta, \quad
U^1 = \sinh \theta \cos  \phi,           \nonumber \\
U^2 &=& \sinh \theta \sin  \phi, \quad
U^3 = \cosh \theta \sinh \eta,
\label{U0123}
\end{eqnarray}
where $\theta$ is the transverse fluid rapidity defined by the formula
\begin{equation}
v_\perp = \sqrt{v_x^2+v_y^2} = \tanh \theta,
\label{thetaperp}
\end{equation} 
$\eta$ is the space-time rapidity~\footnote{For simplicity of notation we use the variables $\theta$ and $\eta$ rather than previously used $\theta_\perp$ and $\eta_\parallel$.}
\begin{eqnarray}
\eta = \frac{1}{2} \ln \frac{t+z}{t-z}, 
\label{etapar} 
\end{eqnarray}
and $\phi$ is the azimuthal angle
\begin{equation}
\phi = \arctan \frac{y}{x}.
\label{phi}
\end{equation}

The scalar function $\theta$ depends on the proper time 
\begin{equation}
\tau = \sqrt{t^2-z^2}
\label{tau}
\end{equation}
and radial distance
\begin{equation}
r = \sqrt{x^2+y^2}.
\label{r}
\end{equation}
It is straightforward to verify in this case that $U$ has a cylindrically symmetric and boost-invariant form, i.e., it satisfies condition $U'^{\mu}(x') = L^{\mu}_{\ \nu} U^{\nu}(x)= U^{\mu}(x')$, where $L^{\mu}_{\ \nu}$ represents a Lorentz boost along the $z$ axis or a rotation about the $z$ axis, and $x' = L \,x$.~\footnote{The boost invariance of vector fields is discussed in more detail in~\CIT{Florkowski:2010zz}.}

In addition to $U^\mu$ we define three other symmetric four-vectors. The first one, $Z^\mu$, defines the longitudinal direction that plays a special role due to the initial geometry of the collision, 
\begin{eqnarray}
Z^0 &=& \sinh \eta_\parallel, \quad
Z^1 = 0, \quad
Z^2 = 0,   \quad
Z^3 = \cosh \eta_\parallel.
\label{Zmu}
\end{eqnarray}
The second four-vector, $R^\mu$, defines a radial direction with respect to the beam,
\begin{eqnarray}
R^0 &=& \sinh \theta \cosh \eta, \quad
R^1 = \cosh \theta \cos  \phi,           \nonumber \\
R^2 &=& \cosh \theta \sin  \phi,           \quad
R^3 = \sinh \theta \sinh \eta,
\label{Rmu}
\end{eqnarray}
while the third four-vector, $\Phi^\mu$, defines the azimuthal direction,
\begin{eqnarray}
\Phi^0 &=& 0, \quad
\Phi^1 = -\sin  \phi,   \quad    
\Phi^2 = \cos  \phi,  \quad        
\Phi^3 = 0.
\label{Phimu}
\end{eqnarray}

The four-vector $U^\mu$ is time-like, while the four-vectors $Z^\mu, X^\mu, Y^\mu$ are space-like. In addition, they are all orthogonal to each other, 
\begin{eqnarray}
U^2 &=& 1, \quad Z^2 = R^2 = \Phi^2 = -1, \nonumber \\
U \cdot Z &=& 0, \quad U \cdot R = 0, \quad U \cdot \Phi = 0, \nonumber \\
Z \cdot R &=& 0, \quad R \cdot \Phi = 0, \quad \Phi \cdot Z = 0.
\label{norm}
\end{eqnarray}

\subsection{Energy--momentum tensor}
\label{sec:enmomten}

In the formalism of hydrodynamics one uses the operator $ \Delta^{\mu \nu} = g^{\mu \nu} - U^\mu U^\nu$, that projects on the three-dimensional space orthogonal to $U^\mu$. It can be shown that 
\begin{equation}
\Delta^{\mu \nu} = -R^\mu R^\nu - \Phi^\mu \Phi^\nu - Z^\mu Z^\nu.
\label{Delta}
\end{equation}
Using Eqs. (\ref{norm}) we find that $Z^\mu, X^\mu$ and $Y^\mu$ are the eigenvectors of $\Delta^{\mu \nu}$,
\begin{equation}
\Delta^{\mu \nu} R_\nu = R^\mu, \quad  \Delta^{\mu \nu} \Phi_\nu = \Phi^\mu, \quad 
\Delta^{\mu \nu} Z_\nu = Z^\mu.
\label{eigen}
\end{equation}

The energy--momentum tensor of the perfect spin fluid has the following structure
\begin{eqnarray}
T^{\mu \nu} = \varepsilon U^\mu U^\nu + P \LB X^\mu X^\nu +  Y^\mu Y^\nu +  Z^\mu Z^\nu \RB. 
\label{enmomten}
\end{eqnarray}
Since we consider boost-invariant and cylindrically symmetric systems, $\varepsilon$ and $P$ (as scalars) may depend only on the (longitudinal) proper time and radial distance, see \EQSTWO{tau}{r}.

In this work, we consider a classical relativistic gas where
\begin{eqnarray}
\varepsilon = \f{g}{2\pi^2} T m^2 \LB e^{\mu/T} + e^{-\mu/T} \RB
\LSB 3 T K_2(z) + m K_1(z)
\RSB
\label{epsilon}
\end{eqnarray}
and
\begin{eqnarray}
P = \f{g}{2\pi^2} T^2 m^2 \LB e^{\mu/T} + e^{-\mu/T} \RB K_2(z).
\label{P}
\end{eqnarray}
Here $K_n(z)$ denotes the modified Bessel function of the second kind the argument $z=m/T$. The two exponential functions correspond to the contributions of the particles and antiparticles (below we use also the notation $\xi = \mu/T$). The factor $g$ denotes the number of internal degrees of freedom. Since our results are independent of $g$, we ignore it in further expressions. In the numerical calculations, we consider two values of the particle mass: $m=$~1 GeV and $m=$~0.3 GeV. We note that spin polarization considered in this work is defined in the particle rest frame, hence, it is well defined form massive particles. Consequently, the limit $m \to 0$ cannot be taken as it is not well defined. 

\begin{figure}[t]
    \centering
    \includegraphics[width=0.55\textwidth]{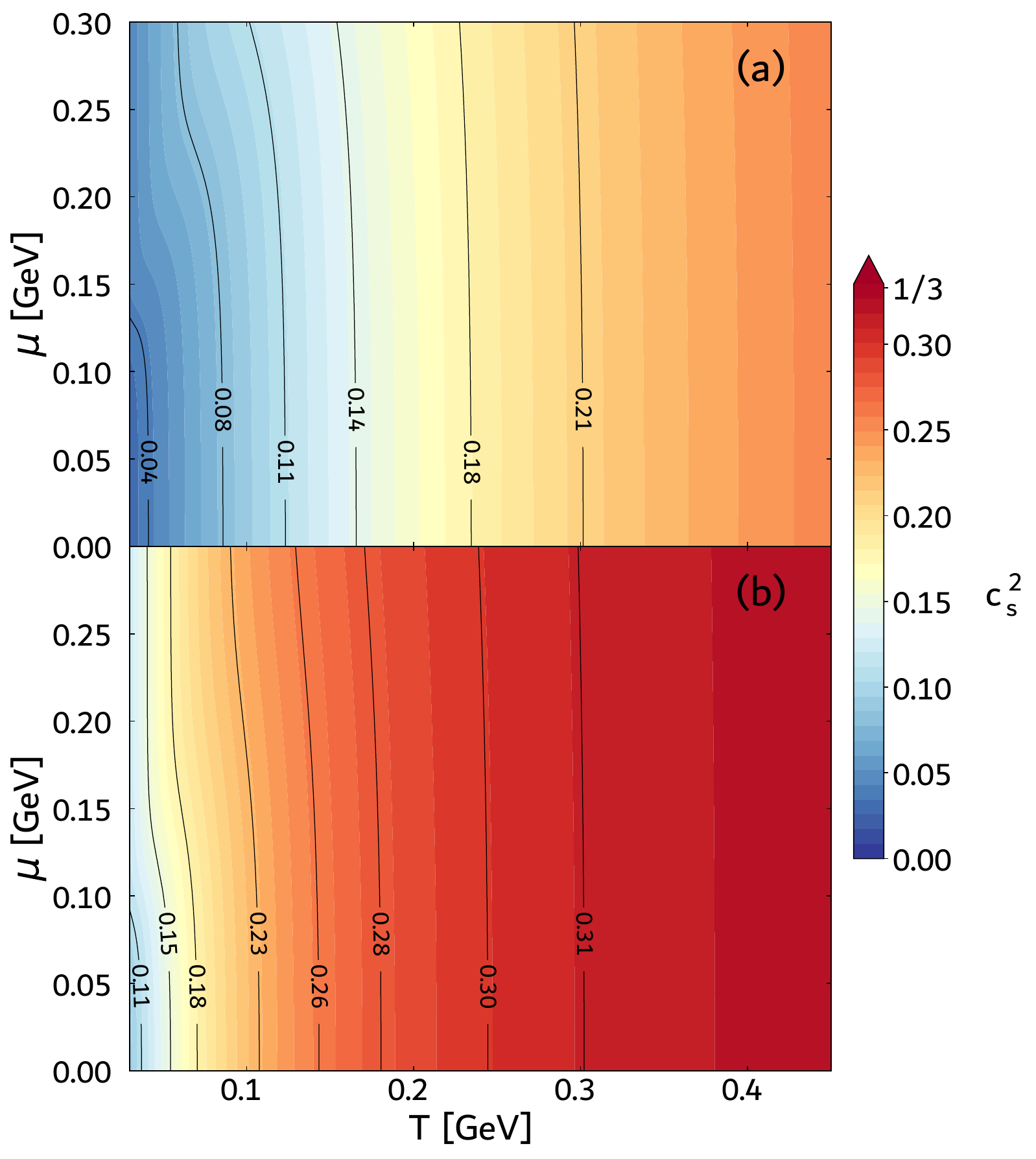}
    \caption{Contour plots of the square of the speed of sound, $c_s^2$, in the temperature and baryon chemical potential $(T, \mu)$ plane. The upper panel (a) shows the results for a particle mass of $m = 1$~GeV, while the lower panel (b) corresponds to a lower mass of $m = 0.3$~GeV. The solid black lines represent contours of constant $c_s^2$.}
    \label{fig:Sound_speed_contours}
\end{figure}

In order to characterize the propagation of perturbations within the fluid, it is essential to determine the speed of sound $c_s$~\CITn{Chojnacki:2007jc}. For a perfect fluid, the square of the speed of sound is defined as the derivative of pressure with respect to energy density at a constant entropy per net baryon $\!s = \sigma/n$,
\begin{equation}
    c_s^2 = \LB \frac{\partial P}{\partial \varepsilon} \RB_{\!s}.
    \label{eq:cs2_def}
\end{equation}
Using standard thermodynamic relations and the properties of Jacobians presented in~\CITn{Landau:1980mil}, this derivative can be expanded in terms of our independent variables $(T, \mu)$ as
\begin{equation}
    c_s^2 = \frac{ \LB\frac{\partial P}{\partial T}\RB_{\!\mu} + \LB\frac{\partial P}{\partial \mu}\RB_{\!T} \LB\frac{\dd\mu}{\dd T}\RB_{\!s} }{ \LB\frac{\partial \varepsilon}{\partial T}\RB_{\!\mu} + \LB\frac{\partial \varepsilon}{\partial \mu}\RB_{\!T} \LB\frac{\dd\mu}{\dd T}\RB_{\!s} },
\end{equation}
where the condition $\dd s = 0$ determines the derivative $(\dd \mu/\dd T)_{\!s}$. 

Using~(\ref{epsilon}) and (\ref{P}), along with the baryon density
\begin{equation}
n = \frac{g}{\pi^2}  T^3 z^2 \sinh(\xi) K_2(z),
\label{eq:ndens}
\end{equation}
and utilizing the recurrence relations for the modified Bessel functions, one can derive an analytical expression for the speed of sound. Introducing a dimensionless auxiliary function $\alpha(z)$,
\begin{equation}
    \alpha(z) \equiv 4 + z \frac{K_1(z)}{K_2(z)},
\end{equation}
the exact square of the speed of sound for a massive gas at finite chemical potential takes the form
\begin{equation}
    c_s^2(z, \xi) = \frac{ \alpha(z)^2 \coth^2\xi - 2\alpha(z)^2 + 5\alpha(z) + z^2 }{ \alpha(z) \left[ (3\alpha(z) + z^2)\coth^2\xi - (\alpha(z)-1)^2 \right] }.
    \label{eq:cs2_full}
\end{equation}
Furthermore, in the regime of vanishing baryon density ($\mu \to 0$, which implies $\xi \to 0$ and \mbox{$\coth^2\xi \to \infty$}), the terms proportional to $\coth^2\xi$ dominate the expression. In this limit~(\ref{eq:cs2_full}) simplifies to the formula for a strictly thermal relativistic massive gas (equivalent to Eq.~(16) of \cite{Castorina:2009de}),
\begin{equation}
    \lim_{\mu \to 0} c_s^2(z) = \frac{\alpha(z)}{3\alpha(z) + z^2} = \frac{4 K_2(z) + z K_1(z)}{(12 + z^2) K_2(z) + 3 z K_1(z)}.
    \label{eq:cs2_limit}
\end{equation}

The contour plots of the function $c_s^2(T,\mu)$ are shown in \FIG{fig:Sound_speed_contours} for two values of the particle mass: $m = 1$~GeV (a) and $m = 0.3$~GeV (b). As expected, for smaller values of the mass, the values of the function $c_s^2(T,\mu)$ are larger. In the limit $m \to 0$,  $c_s^2 \to 1/3$. In the hydrodynamic model, larger values of the speed of sound lead to faster initial cooling and expansion of matter. We will observe this effect in Sec.~\ref{sec:results1}.

\subsection{Spin polarization tensor}
\label{sec:omega}

Following the method used in relativistic magnetohydrodynamics, we decompose the spin polarization tensor into an ``electric'' and a ``magnetic'' part~\footnote{Here we use the original notation from~\CIT{Florkowski:2017ruc}.}
\bea
\omega_{\mu \nu}=k_{\mu} U_{\nu} -k_{\nu} U_{\mu}+t_{\mu\nu} =
k_{\mu} U_{\nu} -k_{\nu} U_{\mu}+\epsilon_{\mu \nu \alpha \beta} U^{\alpha} \omega^{\beta}.
\label{eq:o1}
\eea
The four-vectors $k$ and $\omega$ (the electriclike and magneticlike components, respectively) satisfy the orthogonality conditions
\bea
k_{\mu} U^{\mu} &=& 0, \qquad \omega_{\mu} U^{\mu} = 0.
\eea
Equation~\EQn{eq:o1} implicitly defines the tensor $t_{\mu\nu} = \epsilon_{\mu \nu \alpha \beta} U^{\alpha} \omega^{\beta}$.

Since the four-vectors $k$ and $\omega$ are orthogonal to the flow vector $U$, we can decompose them as follows:
\bea
k^{\mu} = C_{kr} R^{\mu} + C_{k\phi} \Phi^{\mu} 
+ C_{kz} Z^{\mu}, \quad
\omega^{\mu} = C_{\omega r} R^{\mu} + C_{\omega \phi}\Phi^{\mu} + C_{\omega z} Z^{\mu}, 
\eea
where the coefficient functions $C$ depend only on $\tau$ and $r$. For convenience, we introduce a three-vector notation
\bea
\Ckv = (C_{kr},C_{k\phi},C_{kz}), \quad
\Cov = (C_{\omega r}, C_{\omega \phi}, C_{\omega z}),
\eea
A straightforward calculation gives
\beq
\omega_{\mu\nu} &=& C_{k z} (Z_\mu U_\nu - Z_\nu U_\mu)   + C_{k r} (R_\mu U_\nu - R_\nu U_\mu)    + C_{k \phi} (\Phi_\mu U_\nu - \Phi_\nu U_\mu) \nonumber \\
&& + \, \epsilon_{\mu\nu\alpha\beta} U^\alpha (C_{\omega z} Z^\beta + C_{\omega r} R^\beta + C_{\omega \phi} \Phi^\beta),
\label{eq:omegamunu} 
\eeq
and for the dual tensor we find
\beq
{\tilde \omega}_{\mu\nu} &=& C_{\omega z} (Z_\mu U_\nu - Z_\nu U_\mu)   + C_{\omega r} (R_\mu U_\nu - R_\nu U_\mu)    + C_{\omega \phi} (\Phi_\mu U_\nu - \Phi_\nu U_\mu) \nonumber \\
&& - \, \epsilon_{\mu\nu\alpha\beta} U^\alpha (C_{k z} Z^\beta + C_{k r} R^\beta + C_{k \phi} \Phi^\beta).
\label{eq:tomegamunu} 
\eeq
We note that the dula tensor is obtained by the replacements $\Ckv \to \Cov$ and $\Cov \to -\Ckv$.

\subsection{Spin tensor}
\label{sec:spinten}

Following our earlier works, we use the GLW definition of the spin tensor~\CITn{Florkowski:2017ruc}
\begin{equation}
S^{\lambda, \mu \nu} = U^\lambda\big[A(k^\mu U^\nu - k^\nu U^\mu) + A_1 t^{\mu\nu}\big] + \frac{A}{2}\big(t^{\lambda\mu} U^\nu - t^{\lambda \nu} U^\mu + \Delta^{\lambda \mu} k^\nu - \Delta^{\lambda \nu}k^\mu \big),
\label{eq:Sglw}
\end{equation}
where the functions $A$ and $A_1$ are defined by the expressions
\begin{equation}
A = -\frac{4\spin^{2}\cosh{\xi}}{3\pi^{2}}z T^{3}K_{3}(z),
\end{equation}
\begin{equation}
A_1 = \frac{2\spin^{2}\cosh{\xi}}{3\pi^{2}}z T^{3} \left[ z K_{2}(z)+2K_{3}(z)\right].
\end{equation}
Here $\spin^2 = 3/4$ is the eigenvalue of the Casimir operator for SU(2). 

All tensors that have been introduced so far are defined in the GLW pseudogauge. A transition to the canonical spin tensor is made using the formula (see Eq.~(117) in \CITn{Florkowski:2018fap})
\begin{equation}
S^{\lambda, \mu \nu}_{\rm can} = S^{\lambda, \mu \nu} + S^{\mu, \nu \lambda } + S^{\nu, \lambda \mu}.
\label{eq:Scan}
\end{equation}
We note that $S^{\lambda, \mu \nu}_{\rm can}$ defined in this way is totally antisymmetric in all three indices. Substituting \EQn{eq:Sglw} into \EQn{eq:Scan} one finds
\begin{equation}
S^{\lambda, \mu \nu}_{\rm can} = \LB A+A_1 \RB \LB
t^{\lambda\mu} U^\nu - t^{\lambda\nu} U^\mu +
t^{\mu\nu} U^\lambda \RB = \f{{\tilde n}}{3}  \, \epsilon^{\lambda \mu \nu \rho} \, \omega_\rho.
\label{eq:Scan1}
\end{equation}
Interestingly, the spin tensor in the canonical pseudogauge depends only on the magneticlike components of the spin polarization tensor. It has been argued in \CITn{Dey:2023hft} that the canonical spin tensor is preferable to represent spin observables, hence it is directly connected to the generators of rotations. The quantity ${\tilde n}$ in \EQn{eq:Scan1} is the sum of particle and antiparticle densities (${\tilde n} = n/\tanh\xi$, with $n$ given by \EQn{eq:ndens} with $g=2$).

The canonical spin tensor is not conserved, and its divergence equals
\begin{equation}
\p_\lambda S^{\lambda, \mu \nu}_{\rm can} = -\p_\lambda
S^{\mu, \lambda \nu} + \p_\lambda
S^{\nu, \lambda \mu} = 2 T^{[\nu\mu]}_{\rm can},
\label{eq:divScan}
\end{equation}
where the second equality in \EQn{eq:divScan} was derived in~\CITn{Florkowski:2018fap}. It is important to note that \EQn{eq:divScan} is equivalent to the spin conservation in the form $\p_\lambda S^{\lambda, \mu \nu} = 0$. We thus see that the transition between the spin and orbital parts of the total angular momentum may depend on the pseudogauge used (GLW vs. canonical), however, this dependence does not change the real dynamics of the system (equations of motion for $T, \mu, \uv$, and $\omega_{\mu\nu}$).

\section{Spinless hydrodynamic background}
\label{sec:hyd-back}

With dissipation effects neglected and components of the spin polarization tensor included up to the first order, the system of equations of spin hydrodynamics splits into separate equations for a spinless hydrodynamic background and equations that determine the evolution of the spin polarization tensor coupled to this background. Hence, a natural method to deal with such a system is to first solve the standard hydrodynamic equations and then switch to the equation for the spin tensor. In this section, following several earlier works, most importantly \CITn{Baym:1983amj}, we describe our treatment of hydrodynamic equations for cylindrically symmetric and boost-invariant systems.  

\subsection{Baryon current}
\label{sec:Nmu}

For cylindrically symmetric and boost-invariant systems the baryon number conservation law can be cast into the form
\begin{eqnarray}
\left( \frac{\partial}{\partial \tau} 
+ \tanh \theta \frac{\partial}{\partial r} \right) n + n \left[  \left( \frac{1}{\tau} 
+ \frac{\partial \theta}{\partial r} \right) + \tanh \theta \left( 
\frac{1}{r} + \frac{\partial \theta}{\partial \tau} \right) \right] =0.  \label{eq:ncon}
\end{eqnarray}
We recall that the theta parameter indicates transverse rapidity. If the transverse expansion is neglected, $\theta = 0$, from \EQn{eq:ncon} we find that $\dd n/\dd\tau + n/\tau = 0$, with a scaling solution $n = n_0 \tau_0/\tau$, where $n_0$ and $\tau_0$ are integration constants.

\subsection{Energy--momentum tensor}
\label{sec:Tmunu}

Because of the cylindrical symmetry and boost-invariance, the four equations expressing the conservation of energy and linear momentum are reduced to only two equations,
\begin{eqnarray}
\left( \frac{\partial}{\partial \tau} 
+ \tanh \theta \frac{\partial}{\partial r} \right) \varepsilon + (\varepsilon+P) \left[  \left( \frac{1}{\tau} 
+ \frac{\partial \theta}{\partial r} \right) + \tanh \theta \left( 
\frac{1}{r} + \frac{\partial \theta}{\partial \tau} \right) \right] =0  
\label{eq:enmom1}
\end{eqnarray}
and
\begin{eqnarray}
 \left( \tanh \theta \frac{\partial}{\partial \tau} 
+  \frac{\partial}{\partial r} \right) P  + (\varepsilon + P) \left( \tanh \theta \frac{\partial \theta }{\partial r} 
+  \frac{\partial \theta}{\partial \tau} \right)  = 0. \label{eq:enmom2}
\end{eqnarray}
Using the thermodynamic relations $\varepsilon + P = T \sigma + \mu n$ and $\dd\varepsilon = T \dd\sigma + \mu \dd n$ (where $\sigma$ is the entropy density), it can be checked that~\EQ{eq:enmom1} represents the conservation of entropy. On the other hand, \EQ{eq:enmom2} is the relativistic Euler equation that determines the dynamics of the transverse expansion. In the limit $\theta \to 0$ and for transversely uniform systems, \EQ{eq:enmom2} is automatically fulfilled, while ~\EQ{eq:enmom1} is reduced to $\dd\sigma/\dd\tau + \sigma/\tau = 0$, again leading to a scaling solution $\sigma = \sigma_0 \tau_0/\tau$.

Equations \EQn{eq:ncon}, \EQ{eq:enmom1}, and \EQ{eq:enmom2}
are three equations for three unknown functions: $T$, $\mu$, and $\theta$. In the next section we turn to a discussion of their numerical solutions. 

\subsection{Numerical results}
\label{sec:results1}

To solve the equations of the spinless hydrodynamic background, we specify the initial conditions at the proper time $\tau_0 = 1.0$~fm. The initial transverse thermodynamic profiles are initialized proportionally to the Woods--Saxon nuclear density distribution,
\begin{equation}
    \rho(r) = \frac{1}{1 + \exp\LB\frac{r - R}{a}\RB},
\end{equation}
where the nuclear radius $R = 7.0$~fm and the skin thickness $a = 0.54$~fm are chosen to reflect the standard geometric properties of heavy nuclei commonly used in experiments, such as Au or Pb. 

Following this geometric profile, the initial temperature and baryon chemical potential are parameterized as $T(\tau_0, r) = T_0 \, \rho(r)^{1/3} + T_{\rm vac}$  and $\mu(\tau_0, r) = \mu_0 \, \rho(r)$. Here $T_0 = 0.3$~GeV is a central temperature while the parameter $T_{\rm vac} = 0.03$~GeV acts as a small ambient vacuum temperature necessary to ensure numerical stability at large radii where the fluid density vanishes. The central value of the baryon chemical potential is set to $\mu_0 = 0.03$~GeV. Notably, $\mu_0$ is chosen to be exactly an order of magnitude smaller than the central temperature, placing our system in a regime of high temperature and low net-baryon density. The initial transverse fluid rapidity is set to a small constant value $\theta(\tau_0, r) = 0.01$ instead of strictly zero to avoid numerical singularities and ensure a stable initialization of the solver.

Coupled partial differential equations are solved on a one-dimensional radial grid extending up to $r_{\rm max} = 20.0$~fm with a uniform spatial resolution of $\dd r = 0.05$~fm. The temporal evolution is computed up to a final proper time of $\tau_{\rm max} = 10.0$~fm. Spatial gradients are evaluated utilizing second-order finite differences. For the proper-time integration, we employ the Backward Differentiation Formula (BDF).

\begin{figure*}[t]
    \centering
    \includegraphics[width=0.85\textwidth]{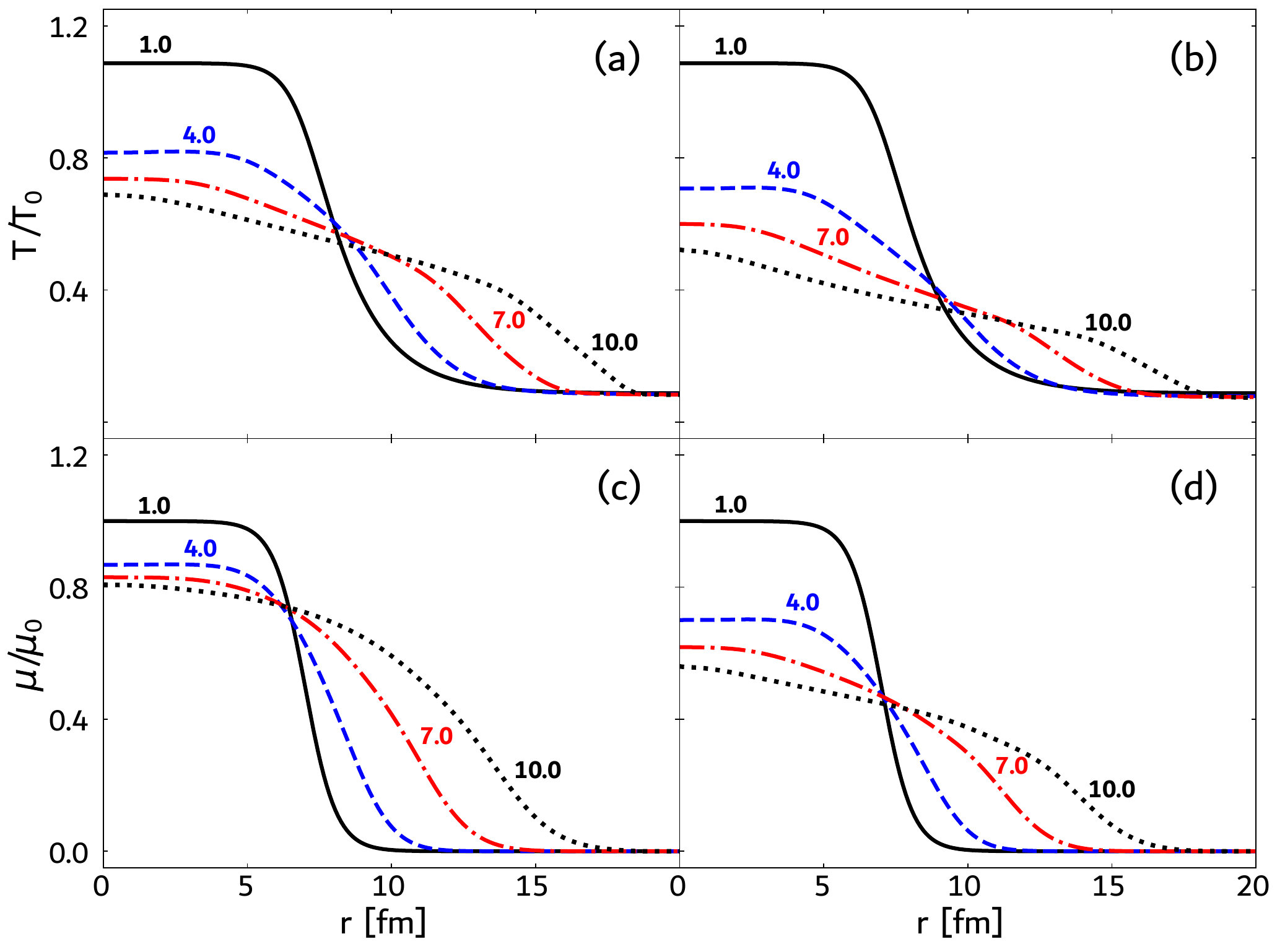}
    \caption{Transverse spatial profiles of the scaled temperature $T/T_0$ (top row) and scaled baryon chemical potential $\mu/\mu_0$ (bottom row) at different proper times $\tau$. The left panels, (a) and (c), display the hydrodynamic evolution for a heavy constituent particle mass of $m = 1$~GeV. The right panels, (b) and (d), correspond to a lighter particle mass of $m = 0.3$~GeV. The initial conditions at $\tau = 1.0$~fm are depicted by the solid black lines. The subsequent expansion stages are shown for $\tau = 4.0$~fm (dashed blue lines), $\tau = 7.0$~fm (dash-dotted red lines), and $\tau = 10.0$~fm (dotted black lines). A central initial temperature $T_0 = 0.3$~GeV and chemical potential $\mu_0 = 0.03$~GeV are used.}
    \label{fig:Hyd_back_comp}
\end{figure*}

To maintain high computational efficiency during the time-stepping process, the thermodynamic quantities and their exact derivatives -- which depend on the modified Bessel functions $K_n(z)$ as defined in Sec.~\ref{sec:enmomten} -- are pre-computed over a fine temperature grid and evaluated continuously using cubic spline interpolation. Additionally, a small numerical viscosity coefficient is introduced via a Laplacian operator to suppress spurious numerical oscillations near steep thermodynamic gradients, without distorting the underlying physical expansion flow.

The results of our numerical calculations are presented in Fig.~\ref{fig:Hyd_back_comp}. Panels (a) and (b) show the spatial profiles of temperature for four values of proper time: $\tau$= 1.0, 4.0, 7.0 and 10.0~fm. Similarly, panels (b) and (d) show the profiles of the baryon chemical potential. The left (right) panels correspond to the case $m = 1$~GeV ($m = 0.3$~GeV). The transverse expansion is faster in the case where the particle mass is smaller. In this case, the speed of sound is larger and the rarefaction wave entering the system is faster.

\subsection{Comparison with earlier results}

To validate our numerical framework, we benchmark our generalized hydrodynamics code against the results obtained in \CITn{Baym:1983amj} for a strictly massless, cylindrically symmetric expanding fluid. Here the initial temperature profile was assumed to directly follow the nuclear density profile, $T(r) \propto \rho(r)$. For a strictly massless gas obeying the Stefan-Boltzmann law ($\varepsilon \propto T^4$), this assumption leads to an initial energy density profile heavily peaked at the center: $\varepsilon(r) \propto \rho(r)^4$. 

\begin{figure}[t]
    \centering
    \includegraphics[width=0.55\textwidth]{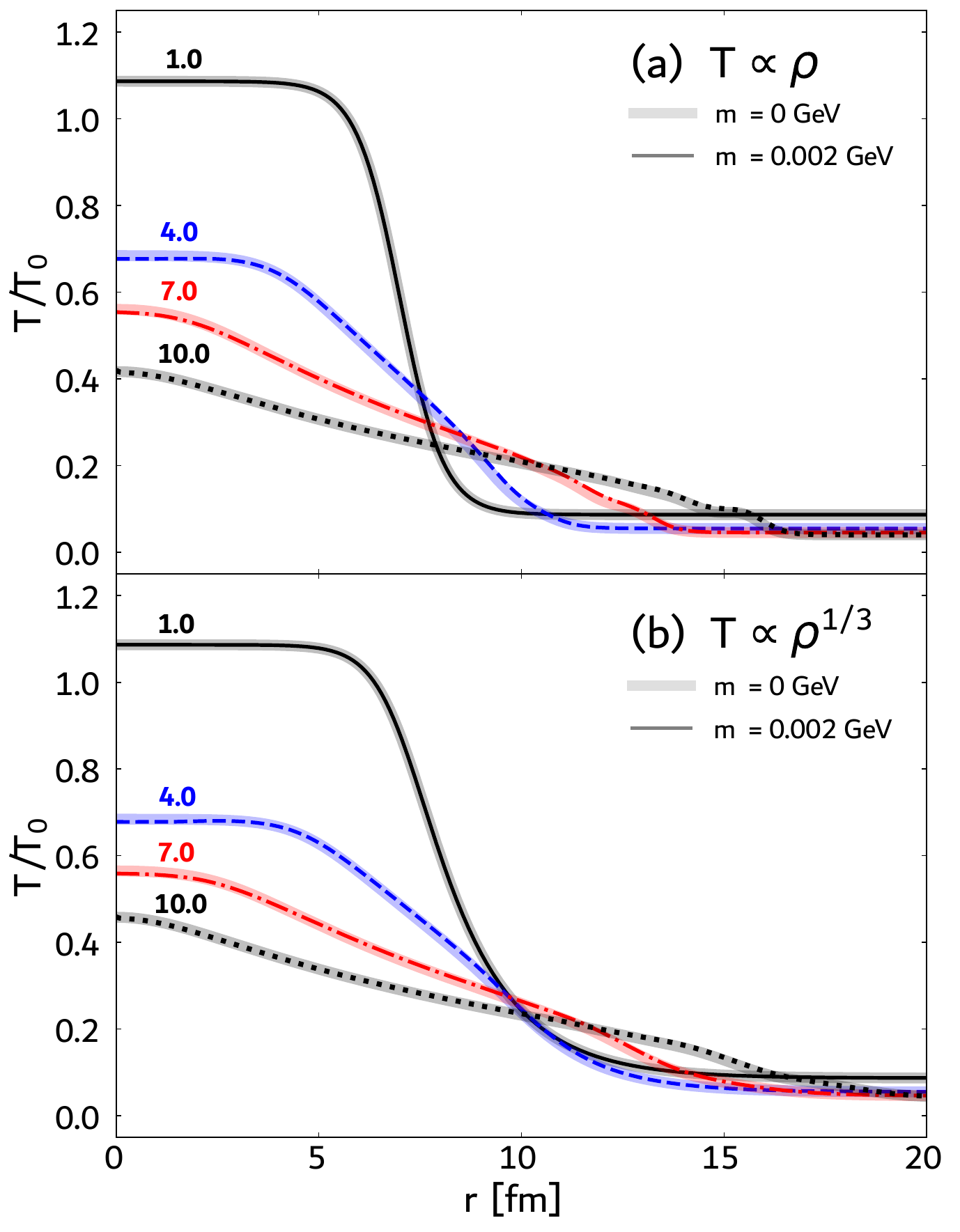}
    \caption{Comparison of the transverse spatial profiles of the scaled temperature $T/T_0$ to validate the massless limit under two distinct initial density scalings: (a) $T \propto \rho$ and (b) $T \propto \rho^{1/3}$ which is in agreement with the wounded nucleon model (WNM). The thick semi-transparent lines represent the solution for a strictly massless gas ($m = 0$~GeV) using the method outlined in \CITn{Baym:1983amj}. The superimposed patterned lines show the numerical results from our massive hydrodynamic solver, evaluated for an exceptionally small particle mass ($m = 0.002$~GeV) and a minimal baryon chemical potential. The initial conditions at $\tau = 1.0$~fm are shown as solid black lines, while the subsequent expansion stages are depicted at proper times $\tau = 4.0$ (dashed blue), $7.0$ (dash-dotted red) and $10.0$~fm (dotted black)..}
    \label{fig:Baym_comp}
\end{figure}

To provide a more physically motivated description, we adopt the wounded nucleon model (WNM)~\CITn{Bialas:1976ed}. This approach is grounded in the experimental observation that the multiplicity of particles in relativistic heavy-ion collisions scales linearly with the number of wounded nucleons. Because the final particle multiplicity is directly related to the total initial entropy of the system, it is physically appealing to scale the initial entropy density $\sigma(r)$, rather than the temperature, with the nuclear density profile, yielding $\sigma(r) \propto \rho(r)$.

Since the entropy density of a massless gas scales as $\sigma \propto T^3$, the WNM implies that the initial temperature profile must follow $T(r) \propto \rho(r)^{1/3}$. Consequently, the initial energy density profile in our validation framework is defined as
\begin{equation}
    \varepsilon(r, \tau_0) = \varepsilon_0 \LSB \frac{\rho(r)}{\rho_0} \RSB^{4/3}.
\end{equation}

While exact formulation presented in \CITn{Baym:1983amj} strictly relies on the equation of state for a massless gas ($\varepsilon = 3P$) with zero net-baryon density, our numerical scheme is fundamentally constructed for a gas of massive particles with a explicitly conserved baryon current. Therefore, testing the massless limit requires a careful numerical setup. To reproduce such a limit within our generalized code, we instantiate our model with an exceptionally small particle mass of $m = 0.002$~GeV. 

To satisfy the requirements of the baryon current conservation equations when testing the massless limit, the initial baryon chemical potential is set to a minimal, nonzero spatial profile $\mu(\tau_0, r) = \mu_0 \rho(r)$, where $\mu_0 = 2 \cdot 10^{-6}$~GeV. Furthermore, a small ambient vacuum temperature $T_{\rm vac}$ is included in the energy density profile to prevent numerical singularities at large radii.

Comparisons between our results and those obtained using the approach of \CIT{Baym:1983amj} are shown in Fig.~\ref{fig:Baym_comp}. We observe a very good agreement between the results obtained with two methods.

\section{Spin densities}
\label{sec:spinden}

The conservation of the spin part of the total angular momentum yields six equations that determine the spacetime dependence of the six independent coefficients of the spin polarization tensor, $\Ckv$ and $\Cov$. For boost invariant systems that are uniform in the transverse plane, these equations decouple and become six independent equations for each coefficient $C$ appearing in the decomposition of the spin polarization tensor (provided that the energy--momentum tensor is included in the leading order in $\omega$). For cylindrically symmetric and boost-invariant systems, the coupling appears between the longitudinal and azimuthal components, which allows us to consider four independent cases denoted below by the symbols {\bf (a)}, {\bf (b)}, {\bf (c)}, and {\bf (d)}.

\noindent
{\bf (a)} Radial electric-like sector defines the dynamics of the coefficient $C_{kr}$ (obtained from the projection $U_\mu R_\nu \p_\lambda S^{\lambda, \mu\nu}=0$),
\begin{equation}
\begin{split}
A \LB \dot{C_{kr}} + {C_{kr}}' \tanh \theta  \RB 
& 
= - C_{kr} \LB \dot{A}  + {A}'  \tanh \theta \RB \\
& - C_{kr}  A \left( \dot{\theta} \tanh \theta + {\theta'}  + \frac{3}{2 \tau}   + \frac{3 \tanh \theta}{2 r} \right) . 
\end{split}
\end{equation}
Here the dot denotes a partial derivative with respect to $\tau$, whereas the prime denotes the partial derivative with respect to the coordinate $r$. A combination ${\dot f} + f^\prime \tanh \theta$ can be interpreted as the convective derivative along the stream lines $\dd f/\dd \tau$.

\noindent 
{\bf (b)} Radial magnetic-like sector, defines the dynamics of $C_{\omega r}$ (from $\Phi_\mu Z_\nu \p_\lambda S^{\lambda, \mu\nu}=0$),
\begin{equation}
\begin{split}
& A_1 \LB \dot{C_{\omega r}}  +  {C_{\omega r}}' \tanh \theta \RB = -C_{\omega r}  \LB   \dot{A_1}  + {A_1}'  \tanh \theta \RB \\
&\quad - C_{\omega r}  \LSB   A_1 \left( \dot{\theta} \tanh \theta + {\theta}' + \frac{1}{\tau} + \frac{ \tanh \theta}{r}  \right) - \frac{A}{2} \left(\frac{1}{\tau} + \frac{\tanh \theta}{r}\right)\RSB.
\end{split}
\end{equation}

\noindent
{\bf (c)} Longitudinal magnetic-like sector, defines the coupled dynamics of $C_{k\phi}$ and $C_{\omega z}$ sector (from the projections $U_\mu \Phi_\nu \p_\lambda S^{\lambda, \mu\nu}=0$ and $R_\mu \Phi_\nu \p_\lambda S^{\lambda, \mu\nu}=0$),
\begin{equation}
\begin{split}
0 &= A \LB \dot{C_{k\phi}}  + {C_{k\phi}}' \tanh \theta \RB - \f{A}{2} \LB  \dot{C_{\omega z}} \tanh \theta+  {C_{\omega z}}'   \RB +  \\
&\quad 
+ C_{k\phi} \LB \dot{A} + {A}' \tanh \theta \RB
- \f{C_{\omega z}}{2} \LB   \dot{A} \tanh \theta +{A}' \RB 
\\
&\quad + A \LSB \f{3}{2} C_{k\phi} \LB  {\theta}'  +  \dot{\theta} \tanh \theta + \frac{1}{\tau} + \frac{2 \tanh \theta}{3 r} \RB    - \f{C_{\omega z}}{2}  \LB  {\theta}' \tanh \theta + \dot{\theta}  + \frac{ \tanh \theta}{\tau} 
\RB \RSB \\
& \quad
+ A_1 C_{\omega z} \left({\theta}' \tanh \theta + \dot{\theta} \right),
\end{split}
\end{equation}

\begin{equation}
\begin{split}\label{eq:rphi}
0&= A_1 \LB \dot{C_{\omega z}}  +  {C_{\omega z}}' \tanh \theta \RB + \f{A}{2} \LB
{C_{k\phi}}'  + \dot{C_{k\phi}} \tanh \theta 
\RB \\
&\quad 
+ \f{C_{k\phi}}{2} \LB \dot{A}  \tanh \theta + {A}'   \RB + C_{\omega z} \LB \dot{A_1}   + {A_1}'  \tanh \theta \RB \\
&\quad + A \LSB  \f{3}{2} C_{k\phi} \LB  {\theta}' \tanh \theta +  \dot{\theta}  + \frac{1}{3 r} \RB - \f{C_{\omega z}}{2} \LB  {\theta}'  +  \dot{\theta} \tanh \theta + \frac{ \tanh \theta}{r} \RB \RSB \\
&\quad  +  A_1 \, C_{\omega z} \left( {\theta}'  +  \dot{\theta} \tanh \theta + \frac{1}{\tau} + \frac{ \tanh \theta}{r}   \right).
\end{split}
\end{equation}

\noindent
{\bf (d)} Longitudinal electric-like sector defines the coupled dynamics of $C_{kz}$ and $C_{\omega \phi}$ (obtained from the projections $U_\mu Z_\nu \p_\lambda S^{\lambda, \mu\nu}=0$ and $Z_\mu R_\nu \p_\lambda S^{\lambda, \mu\nu}=0$)

\begin{equation}
\begin{split}\label{eq:uz}
0 &= A \LB \dot{C_{kz}} + {C_{kz}}' \tanh \theta \RB + \f{A}{2} \LB \dot{C_{\omega \phi}} \tanh \theta + {C_{\omega \phi}}'  \RB\\
&\quad + C_{kz} \LB \dot{A} + {A}' \tanh \theta \RB 
+ \f{C_{\omega \phi} }{2} \LB \dot{A} \tanh \theta + {A}' \RB
\\
&\quad + A \LSB \frac{3}{2} \, C_{kz} \LB   {\theta}'  +  \dot{\theta} \tanh \theta + \frac{ 2}{3 \tau} + \frac{\tanh \theta}{r} \RB + \frac{C_{\omega \phi}}{2} \LB  {\theta}' \tanh \theta +  \dot{\theta}  + \frac{1}{r} \RB  \RSB \\
&\quad  - A_1 \, C_{\omega \phi} \left({\theta}' \tanh \theta + \dot{\theta} \right),
\end{split}
\end{equation}

\begin{equation}
\begin{split}
0 &=  A_1 \LB \dot{C_{\omega \phi}}  + {C_{\omega \phi}}' \tanh \theta  \RB - \frac{A}{2}
\LB \dot{C_{kz}} \tanh \theta + {C_{kz}}'  \RB\\
&\quad 
+ C_{\omega \phi} \LB \dot{A_1}  + 
{A_1}' \tanh \theta \RB
- \frac{C_{kz}}{2} \LB \dot{A}  \tanh \theta + {A}' \RB 
 \\
&\quad 
- A \LSB \frac{3}{2} C_{kz} \LB  {\theta}' \tanh \theta +  \dot{\theta} + \frac{\tanh \theta}{3 \tau} \RB + \frac{C_{\omega \phi}}{2} \LB  {\theta}'  + \dot{\theta} \tanh \theta + \frac{1}{\tau} \RB \RSB \\
&\quad 
 + A_1 \, C_{\omega \phi} \left( {\theta}'  + \dot{\theta} \tanh \theta + \frac{1 }{\tau} + \frac{ \tanh \theta}{r} \right).
\end{split}
\end{equation}

The coupling between the $C_{k z}$ and $C_{\omega \phi}$ components, as in {\bf (d)}, and between the $C_{\omega z}$ and $C_{k \phi}$ components, as in {\bf (c)}, was previously found in the analysis of the Gubser flow~\CITn{Singh:2020rht, Singh:2026wvf}. However, our finding is more general because we do not assume conformal symmetry used in the Gubser flow. Our initial profiles of $\Ckv$ and $\Cov$ in the transverse plane are arbitrary.

It is important to note that in cases {\bf (c)} and {\bf (d)} the terms $C_{\omega \phi}/r$ and $C_{k \phi}/r$ appear. This indicates that the coefficients $C_{\omega \phi}(\tau,r)$ and $C_{k \phi}(\tau,r)$ should vanish for $r=0$. This is a natural assumption, since the azimuthal components of vectors cannot be finite at the origin of the polar coordinates. As a consequence, in the numerical calculations we always assume the initial conditions $C_{\omega \phi}(\tau_0,r=0)=0$ and $C_{k \phi}(\tau_0,r=0)=0$. It turns out that the hydrodynamic evolution preserves these conditions at later times. 

\begin{figure*}[t]
    \centering
    \includegraphics[width= 0.85\textwidth]{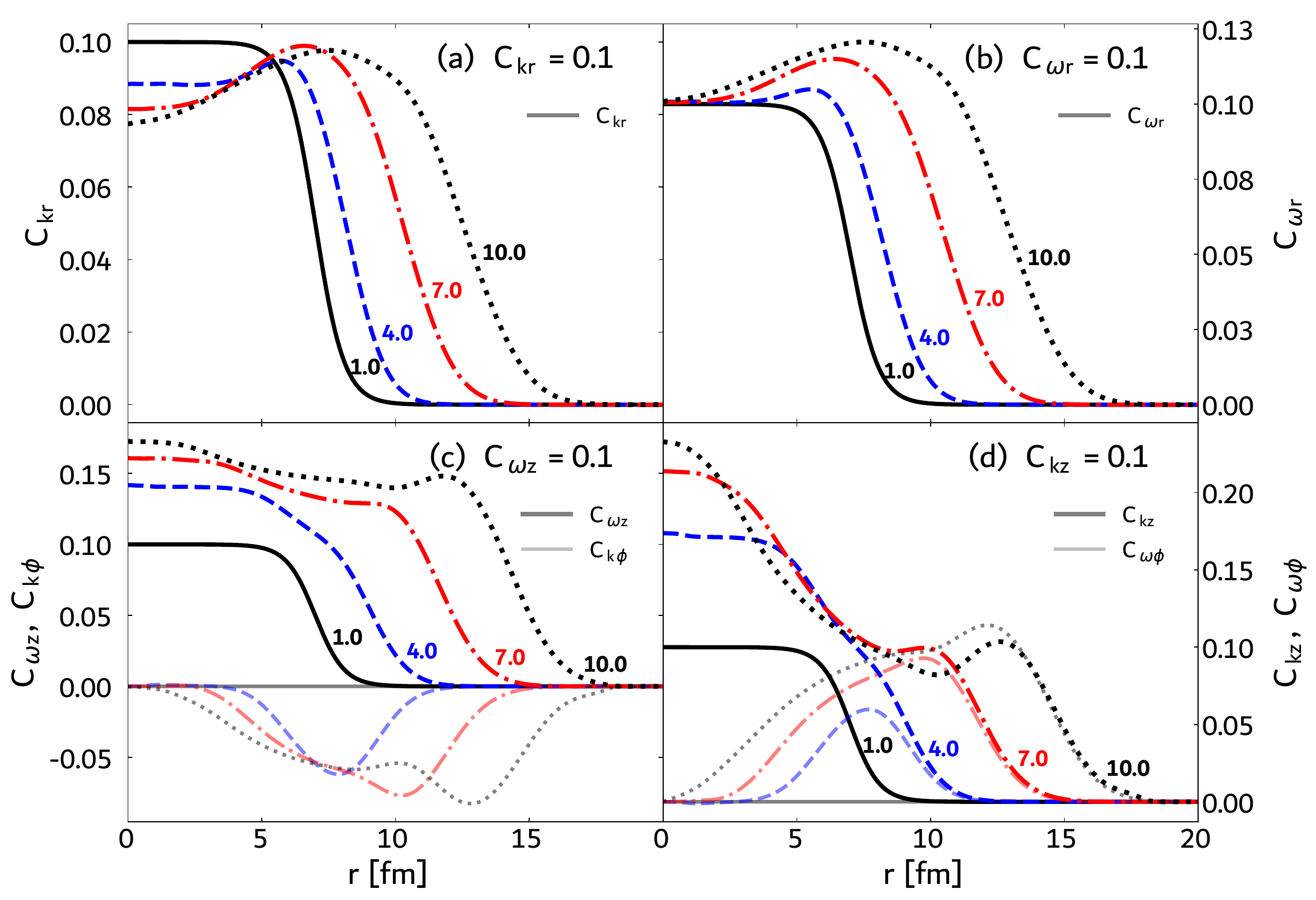}
    \caption{Time evolution of the spin polarization tensor components as a function of the radial distance $r$ for a heavy particle mass of $m = 1$~GeV. The initial conditions at $\tau = 1.0$~fm are defined by a modified Woods--Saxon profile (solid black lines). Subsequent expansion stages are shown for $\tau = 4.0$~fm (dashed blue lines), $7.0$~fm (dash-dotted red lines), and $10.0$~fm (dotted black lines). Panels (a) and (b) demonstrate the independent evolution of the radial components $C_{kr}$ and $C_{\omega r}$. Panels (c) and (d) present the coupled dynamics: initializing $C_{\omega z}$ induces $C_{k\phi}$, and initializing $C_{kz}$ induces $C_{\omega \phi}$. The initialized components are plotted with thick lines, while the dynamically induced components are shown with semi-transparent, thinner lines.}
    \label{fig:Spin_1GeV}
\end{figure*}

\begin{figure*}[htbp]
    \centering
    \includegraphics[width= 0.85\textwidth]{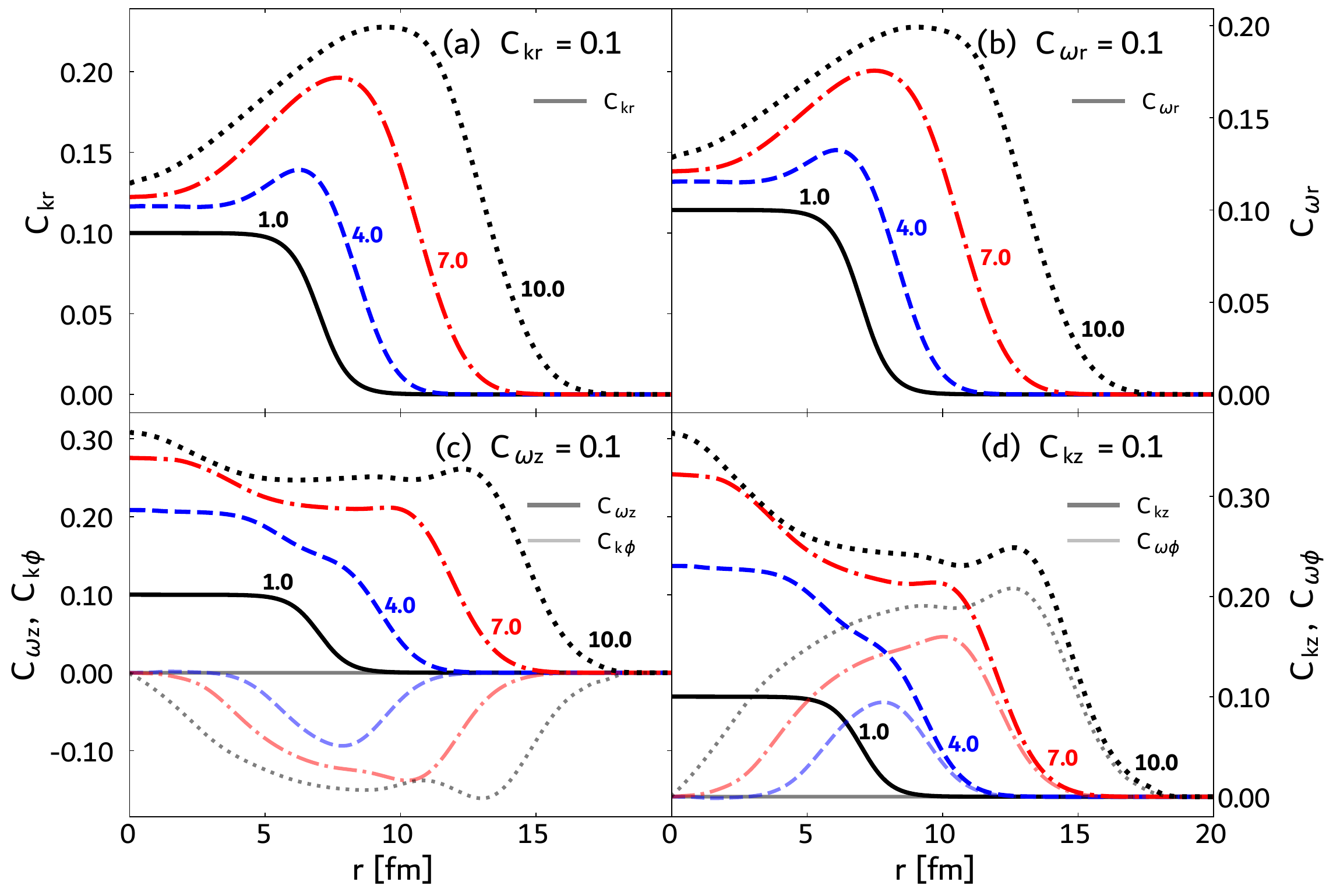}
    \caption{Time evolution of the spin polarization tensor components as a function of the radial distance $r$, evaluated for a lighter particle mass of $m = 0.3$~GeV. The layout, initial conditions, and line styles correspond exactly to those in \FIG{fig:Spin_1GeV}.}
    \label{fig:Spin_0p3GeV}
\end{figure*}

As explained above, to solve the spin evolution equations numerically, we decouple the problem by first evaluating the spinless hydrodynamic background and then solving the spin dynamics using the pre-calculated background variables ($T$, $\mu$, and $\theta$) as interpolating functions. The solution method was adapted from the hydrodynamic background. The spatial derivatives are computed using finite differences on a one-dimensional radial grid, while the time integration is performed using the BDF. To suppress spurious numerical oscillations near steep thermodynamic gradients without altering the physical dynamics, a small numerical viscosity term is explicitly added to the evolution equations.

To illustrate the dynamics of the spin polarization tensor, we initialize the system by setting only one specific component to a nonzero value at the initial proper time $\tau_0 = 1.0$~fm, while keeping all other components strictly zero. The initial nonzero spin density profiles are assumed to be proportional to the nuclear density distribution, $C_i(\tau_0, r) = 0.1 \rho(r)$. Based on the structure of the evolution equations introduced above, we distinguish four independent cases, which are presented in panels (a)--(d) of \FIG{fig:Spin_1GeV} and \FIG{fig:Spin_0p3GeV}. In panels (a) and (b), we initialize the radial electric-like ($C_{kr}$) and magnetic-like ($C_{\omega r}$) components, respectively. Their evolution is completely decoupled from the rest of the spin tensor, expanding radially without inducing any other spin components. In contrast, panels (c) and (d) demonstrate coupled dynamics. Initializing the longitudinal magnetic-like component $C_{\omega z}$ dynamically induces the azimuthal electric-like component $C_{k \phi}$ (case c), whereas initializing the longitudinal electric-like component $C_{k z}$ leads to the dynamic generation of the azimuthal magnetic-like component $C_{\omega \phi}$ (case d).

Figures~\ref{fig:Spin_1GeV} and \ref{fig:Spin_0p3GeV} correspond to the cases $m = 1$~GeV and $m = 0.3$~GeV, respectively. The mass difference affects mainly the evolution of the coefficients $C_{k r}$ whose central value decreases for $m = 1$~GeV and increases for $m = 0.3$~GeV. In other cases, we observe a moderate increase of the coefficients with time, and the outwards shift of the maximum, an effect caused by the transverse expansion. 
An increase of the coefficients $C_i$ in time has been observed previously in perfect spin hydrodynamics in the boost-invariant transversely homogeneous geometry
\cite{Drogosz:2024lkx, Drogosz:2026qbo}.
It is likely that an inclusion of dissipation would change this picture and cause those coefficients to decrease. This may be verified by future studies of dissipative spin hydrodynamics.

In the coupled channels, we find some interesting features: the increase of the longitudinal magnetic component $C_{\omega z}$ produces a negative azimuthal electric component $C_{k \phi}$, while the increase of the longitudinal electric component $C_{\omega z}$ produces a positive azimuthal magnetic component $C_{\omega \phi}$. This behavior shares some similarities with the Maxwell equations.

\section{Pauli--Lubański vector}
\label{sec:PLvector}

In this section, we calculate the spin polarization of particles described by our hydrodynamic model. We assume that at some fixed freeze-out temperature, the system undergoes a transition from fluid-like behavior to free streaming. The mean spin polarization of the particles with four-momentum $\pvv$ is obtained from the Cooper--Frye formula~\CITn{Becattini:2013fla, Florkowski:2018fap}
\beq
\langle  \pi_{\mu} \rangle =\frac{E_p\frac{\dd\Pi _{\mu }(p)}{\dd^3 p}}{E_p\frac{\dd{\cal{N}}(p)}{\dd^3 p}},
\label{averagePL}
\eeq
where the numerator defines the momentum density of the Pauli--Lubański vector~\CITn{Becattini:2013fla, Florkowski:2018fap}
\begin{equation}
E_p\frac{\dd\Pi _{\mu }(p)}{\dd^3 p} = -\f{ \cosh(\xi)}{(2 \pi )^3 m}
\int \dd \Sigma _{\lambda } \, p^{\lambda } \,
e^{-\beta \cdot p} \,
\tilde{\omega }_{\mu \beta }p^{\beta }, \label{PDPLV}
\end{equation}
while the denominator gives the momentum density of particles
\beq
E_p\frac{\dd{\cal{N}}(p)}{\dd^3 p}&=&
\f{4 \cosh(\xi)}{(2 \pi )^3}
\int \dd \Sigma_{\lambda } \, p^{\lambda } 
\,
e^{-\beta \cdot p} .
\label{densityofpart}
\eeq
Here, we include both particles and antiparticles, which is reflected in the common prefactor $\cosh\xi$. However, this factor is canceled in the expression for the mean spin polarization (as a consequence of using the classical statistics)
\beq
\langle  \pi_{\mu} \rangle = -\f{1}{4m}
\frac{\int \dd \Sigma _{\lambda } \, p^{\lambda } \,
e^{-\beta \cdot p} \tilde{\omega }_{\mu \beta }p^{\beta }}
{\int \dd \Sigma _{\lambda } \, p^{\lambda } \,
e^{-\beta \cdot p} }.
\label{averagePL2}
\eeq
We note that the formula \EQn{PDPLV} is the same for the GLW and canonical versions of the spin tensor, as demonstrated in~\CITn{Florkowski:2017dyn}. The expression ${\tilde \omega}_{\mu\nu} p^\nu$ appearing in \EQn{PDPLV} can be written as
\beq
{\tilde \omega}_{\mu\nu} p^\nu &=& 
\LB C_{\omega z} Z_\mu + C_{\omega r} R_\mu + C_{\omega \phi} \Phi_\mu \RB \, U \cdot p - U_\mu \, \LB C_{\omega z}  Z \cdot p  + C_{\omega r} R \cdot p  
 + C_{\omega \phi}  \, \Phi \cdot p \RB  \nonumber \\
&& - \, \epsilon_{\mu\nu\alpha\beta} \, p^\nu U^\alpha (C_{k z} Z^\beta + C_{k r} R^\beta + C_{k \phi} \Phi^\beta).
\label{eq:tomegamunupnu} 
\eeq
The explicit components of this vector are given in Appendix~\ref{app:PL}.
%

\subsection{Freeze-out hypersurface}

The hypersurface $\Sigma$ invariant under boosts and cylindrically symmetric is parameterized with the help of three coordinates: the azimuthal angle $\phi$, the spacetime rapidity $\eta$, and an additional variable $\zeta$~\CITn{Florkowski:2010zz},
\begin{equation}
\Sigma^\mu = \LB \tau(\zeta) \cosh\eta, r(\zeta) \cos\phi, r(\zeta) \sin\phi, \tau(\zeta) \sinh\eta \RB. 
\label{Sigma}
\end{equation}
Here $\zeta$ is used to parametrize the freeze-out hypersurface (strictly speaking, its projection) in the $\tau-r$ plane. A standard choice is to determine the curve $\LB \tau(\zeta), r(\zeta) \RB$ by the condition of constant temperature, $T(\tau,\zeta)$~=~const.

The volume element of the hypersurface~\EQn{Sigma} is then~\CITn{Florkowski:2010zz}
\begin{equation}
\dd\Sigma^\mu = \LB \f{\dd r(\zeta)}{\dd\zeta} \cosh\eta, \f{\dd\tau(\zeta)}{\dd\zeta} \cos\phi, \f{\dd\tau(\zeta)}{\dd\zeta} \sin\phi, \f{\dd r(\zeta)}{\dd\zeta} \sinh\eta \RB r(\zeta) \tau(\zeta) \dd\zeta \dd\eta \dd\phi. 
\label{dSigma}
\end{equation}
With the standard momentum parameterization in terms of rapidity $y$ and transverse momentum $p_T$ (the transverse mass $m_T = \sqrt{m^2 + p_T^2}$)
\begin{equation}
p^\mu = \LB m_T \cosh(y), p_T \cos(\phi_p), p_T \sin(\phi_p), m_T \sinh(y) \RB
\end{equation}
we find

\beq
p \cdot \dd\Sigma = \LSB m_T \cosh(y-\eta_\parallel) \f{\dd r}{\dd\zeta} - p_T \cos(\phi-\phi_p) \f{\dd\tau}{\dd\zeta} \RSB
r(\zeta) \tau(\zeta) \dd\zeta \dd\eta \dd\phi .
\label{eq:pSigma}
\eeq
We also have
\beq
p \cdot u = m_T \cosh(\theta) \cosh(y-\eta) - p_T \sinh(\theta) \cos(\phi-\phi_p).
\label{eq:pu}
\eeq
Equations \EQn{eq:pSigma} and \EQn{eq:pu} can be used in \EQn{averagePL2} to obtain the Pauli--Lubański four-vector.

\begin{figure}[t]
    \centering
    \includegraphics[width=0.65\textwidth]{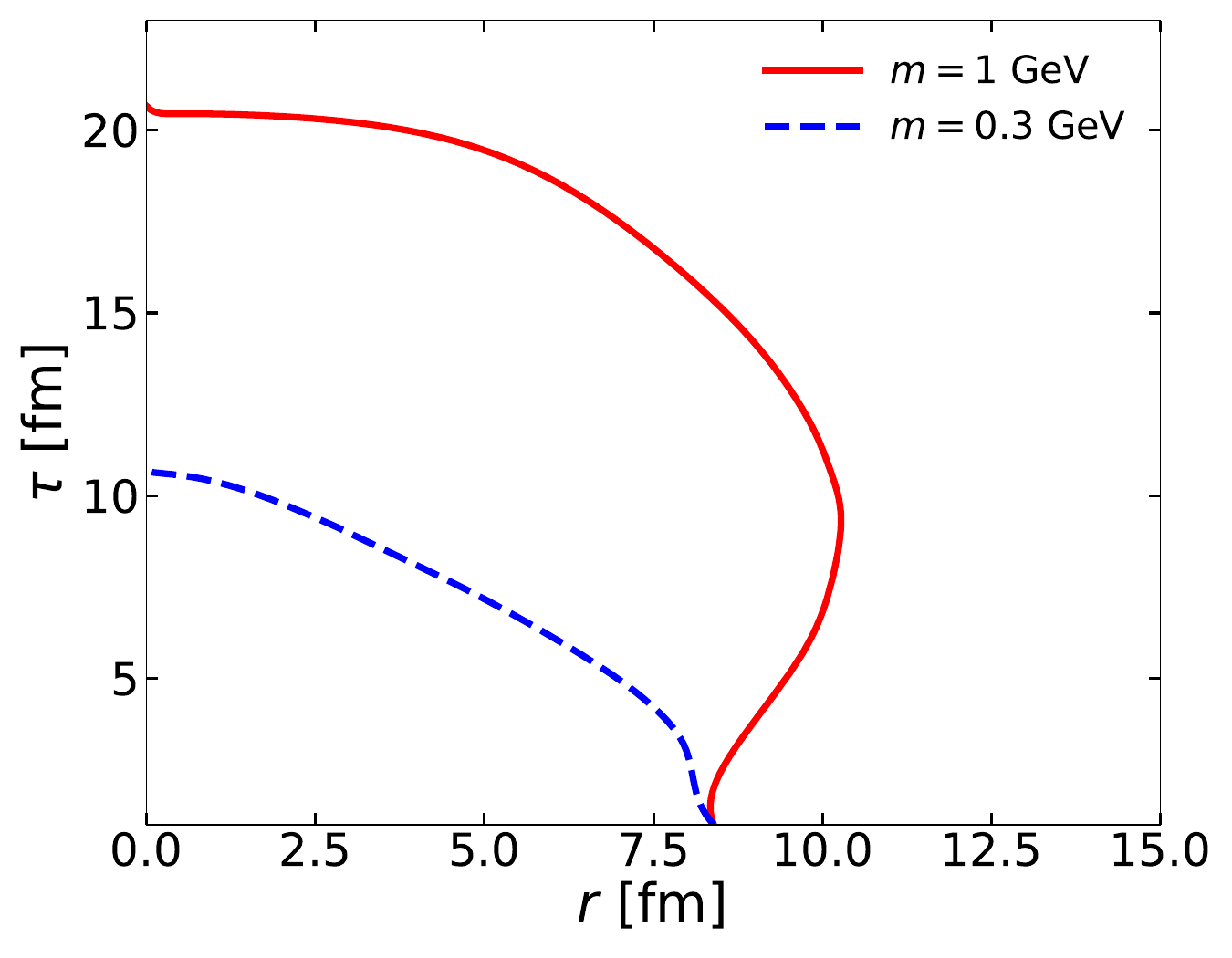}
    \caption{Projection ($\eta =0$ and $\phi=0$) of the isothermal freeze-out hypersurfaces corresponding to a constant temperature of $T = 0.15$~GeV. The solid black line indicates the expansion profile for a fluid composed of massive particles with $m = 1$~GeV, while the dashed line represents the lower mass case of $m = 0.3$~GeV. The curves display the proper time $\tau$ as a function of the radial coordinate $r$.}
    \label{fig:freezeout}
\end{figure}

\subsection{Hypersurface integrals}

In the case of Boltzmann statistics, the distributions function $e^{-\beta p \cdot u}$ factorizes into a product of functions depending on $\cosh(y-\eta)$ and $\cos(\phi-\phi_p)$. As a consequence, one can analytically perform the integrals over spacetime  rapidity $\eta$ and azimuthal angle $\phi$. The first integrals can be expressed by the modified Bessel function of the second kind $K_n(m/T)$, while the second integrals yieled by the modified Bessel function of the first kind $I_n(m/T)$. 

For example, the integral in the numerator of \EQ{averagePL2} has a structure
\beq
\int \dd \Sigma _{\lambda } \, p^{\lambda } \,
e^{-\beta \cdot p} \,
\tilde{\omega }_{\mu \beta }p^{\beta }  = \int
\LB m_T F_\mu \f{\dd r}{\dd\zeta} - p_T G_\mu \f{\dd\tau}{\dd\zeta} \RB
r(\zeta) \tau(\zeta) \dd\zeta,
\label{FG}
\eeq
where (for $y=0$) the four-vectors $F_\mu$ and $G_\mu$ are defined by the integrals
\beq
F_\mu = \int \cosh(\eta) 
\exp\LSB - \f{m_T}{T} \cosh(\theta) \cosh(\eta) \RSB \LSB \int 
\exp\LSB  \f{p_T}{T} \sinh(\theta) \cos(\phi-\phi_p)\RSB \,
\tilde{\omega }_{\mu \beta }p^{\beta } \dd\phi  \RSB \dd\eta
\eeq
and
\beq
\!\!\! G_\mu = \!\!\! \int 
\exp\LSB - \f{m_T}{T} \cosh(\theta) \cosh(\eta) \RSB \LSB \int \cos(\phi\!-\!\phi_p) 
\exp\LSB  \f{p_T}{T} \sinh(\theta) \cos(\phi\!-\!\phi_p)\RSB \,
\tilde{\omega }_{\mu \beta }p^{\beta } \dd\phi  \RSB \dd\eta .
\eeq
Let us introduce a shorthand notation
\beq\label{eq:gothic}
\mgoth = \frac{m_T \cosh \theta}{T}, \qquad \pgoth = \frac{p_T \sinh \theta}{T}.
\eeq
The integrals over $\eta$ and $\phi$ give
\beq
F_0 &=& 2 \pi  C_{\omega r} p_T \, \kappa \,
I_1(\pgoth), \nn \\
F_1 &=& 4 \pi  p_T \sin (\phi_p) (C_{k z} \cosh \theta-C_{\omega \phi} \sinh \theta) K_1 (\mgoth)
I_0(\pgoth) 
\nn \\
&& -2 \pi  m_T \, \kappa \, 
I_1(\pgoth) (C_{k z} \sinh \theta \sin (\phi_p)
-C_{\omega \phi} \cosh (\theta) \sin (\phi_p)+C_{\omega r} \cos (\phi_p)), \nn \\
F_2 &= & 4 \pi  p_T \cos (\phi_p) (C_{\omega \phi} \sinh \theta-C_{k z} \cosh \theta) K_1 (\mgoth)
   I_0(\pgoth) \nn \\
&& -2 \pi  m_T \, \kappa \, I_1(\pgoth) (-C_{k z} \sinh \theta \cos (\phi_p)+C_{\omega \phi} \cosh
   \theta \cos (\phi_p)+C_{\omega r} \sin (\phi_p)), \nn \\
F_3 &=& 2 \pi  p_T (C_{k \phi} \cosh \theta+C_{\omega z} \sinh \theta) \, \kappa \,I_1(\pgoth) \nn \\
&& -4 \pi  m_T (C_{k \phi} \sinh \theta+C_{\omega z} \cosh (\theta
   )) K_1(\mgoth) I_0(\pgoth).
\eeq
where we have put
\beq
\kappa = K_0 (\mgoth) + K_2 (\mgoth).
\eeq
One can explicitly check that $p^\mu F_\mu = 0$. Similarly, the components of the four-vector $G_\mu$ are
\beq
G_0 &=& 4 \pi  C_{\omega r} K_1(\mgoth) \LSB p_T I_2(\pgoth) +
T \csch (\theta) I_1(\pgoth) \RSB, \nn \\
G_1 &=& \f{4 \pi}{p_T} I_1(\pgoth) \LSB p_T^2 \sin (\phi_p) (C_{k z} \cosh \theta-C_{\omega \phi} \sinh \theta)
   K_0(\mgoth) \right. \nn \\
&& \left. -m_T T K_1(\mgoth) (\sin (\phi_p)
   (C_{k z}-C_{\omega \phi} \coth \theta)+C_{\omega r} \csch(\theta) \cos (\phi_p))\RSB \nn \\
&& -4 \pi  m_T
   K_1(\mgoth) I_2(\pgoth) \LSB C_{k z} \sinh \theta \sin (\text{$\phi
   $p})-C_{\omega \phi} \cosh \theta \sin (\phi_p)+C_{\omega r} \cos (\phi_p) \RSB, \nn \\
G_2 &=& \frac{4 \pi}{p_T}  I_1(\pgoth) \LSB  p_T^2 \cos (\phi_p) (C_{\omega \phi} \sinh
   \theta-C_{k z} \cosh \theta) K_0(\mgoth) 
   \right. \nn \\
&& \left.
+ m_T T K_1(\mgoth) (\cos (\phi_p)
   (C_{k z}-C_{\omega \phi} \coth \theta)-C_{\omega r} \csch(\theta) \sin (\phi_p)) \RSB \nn \\
&& -4 \pi  m_T K_1(\mgoth) I_2(\pgoth) \LSB -C_{k z} \sinh \theta \cos (\phi_p)+C_{\omega \phi} \cosh \theta \cos (\phi_p)+C_{\omega r} \sin (\phi_p) \RSB, 
\nn \\
G_3 &=& 4 \pi  \sinh \theta \LSB (C_{k \phi} \coth \theta+C_{\omega z}) K_1(\mgoth) \LSB p_T
   I_2(\pgoth)+T \csch(\theta) I_1(\pgoth) \RSB
\right. \nn \\
&& \left. -m_T
   (C_{k \phi}+C_{\omega z} \coth \theta) K_0(\mgoth) I_1(\pgoth) \RSB.
\eeq
In this case we can also verify that $p^\mu G_\mu =0$.

To evaluate the mean spin polarization vector from the Cooper--Frye formalism, we dynamically extract the relevant isothermal freeze-out contour from the background hydrodynamic evolution (see \FIG{fig:freezeout}). The space-time profiles of the background thermodynamic variables, the fluid velocity, and all the spin polarization tensor components are mapped onto this one-dimensional curve utilizing a high-resolution regular grid interpolation. To ensure numerical stability during the integration, appropriate safeguards are implemented near the origin ($r \to 0$), where L'Hôpital's rule is applied to properly evaluate divergent-looking geometric source terms.

The analytical hypersurface integrands of the momentum density of the Pauli--Lubański vector and the momentum density of particles \EQSTWO{PDPLV}{densityofpart} -- which explicitly depend on the modified Bessel functions $K_n$ and $I_n$ -- are evaluated at each point along the extracted curve. This evaluation is performed over a uniformly discretized grid of transverse momenta $(p_x, p_y)$ spanning the range from $-4.0$ to $4.0$ GeV. The continuous integration over the freeze-out hypersurface is carried out numerically via a discrete summation, strictly accounting for the appropriate cylindrical volume measure. The mean Pauli--Lubański four-vector is then obtained by \EQ{averagePL}.

To ensure the accuracy of this semi-analytical approach, we independently cross-checked our results using a fully numerical integration. Instead of utilizing the analytically pre-integrated formulas with Bessel functions, the integrals in \EQSTWO{PDPLV}{densityofpart} over the spacetime rapidity $\eta$ and the azimuthal angle $\phi$ were evaluated explicitly over a highly resolved, two-dimensional discrete grid: $\eta \in [-4.0, 4.0]$ and $\phi \in [0, 2\pi)$. In this alternative scheme, the covariant components of the Pauli--Lubański vector were computed directly from their definitions. This involved explicitly constructing the local basis four-vectors at each integration node, calculating the exact four-vector scalar products, and dynamically evaluating the local Boltzmann weights. This fully numerical integration scheme yielded exact agreement with the semi-analytical method, providing a validation of our computational framework.

\subsection{Boost to the particle rest frame}

Finally, the mean spin four-vector computed in the laboratory frame, $\langle \pi^\mu_{\rm LAB} \rangle$, must be transformed to the particle rest frame (PRF) to yield the physical, observable polarization components. In our numerical implementation, this transformation is executed point-by-point over the discrete two-dimensional transverse momentum grid $(p_x, p_y)$. For each grid node, we evaluate the transverse momentum magnitude $p_T = \sqrt{p_x^2 + p_y^2}$ and the corresponding transverse mass $m_T = \sqrt{m^2 + p_T^2}$, which acts as the effective energy of the particle in the transverse plane.

\begin{figure*}[h!]
    \centering
    \includegraphics[width=0.95\textwidth]{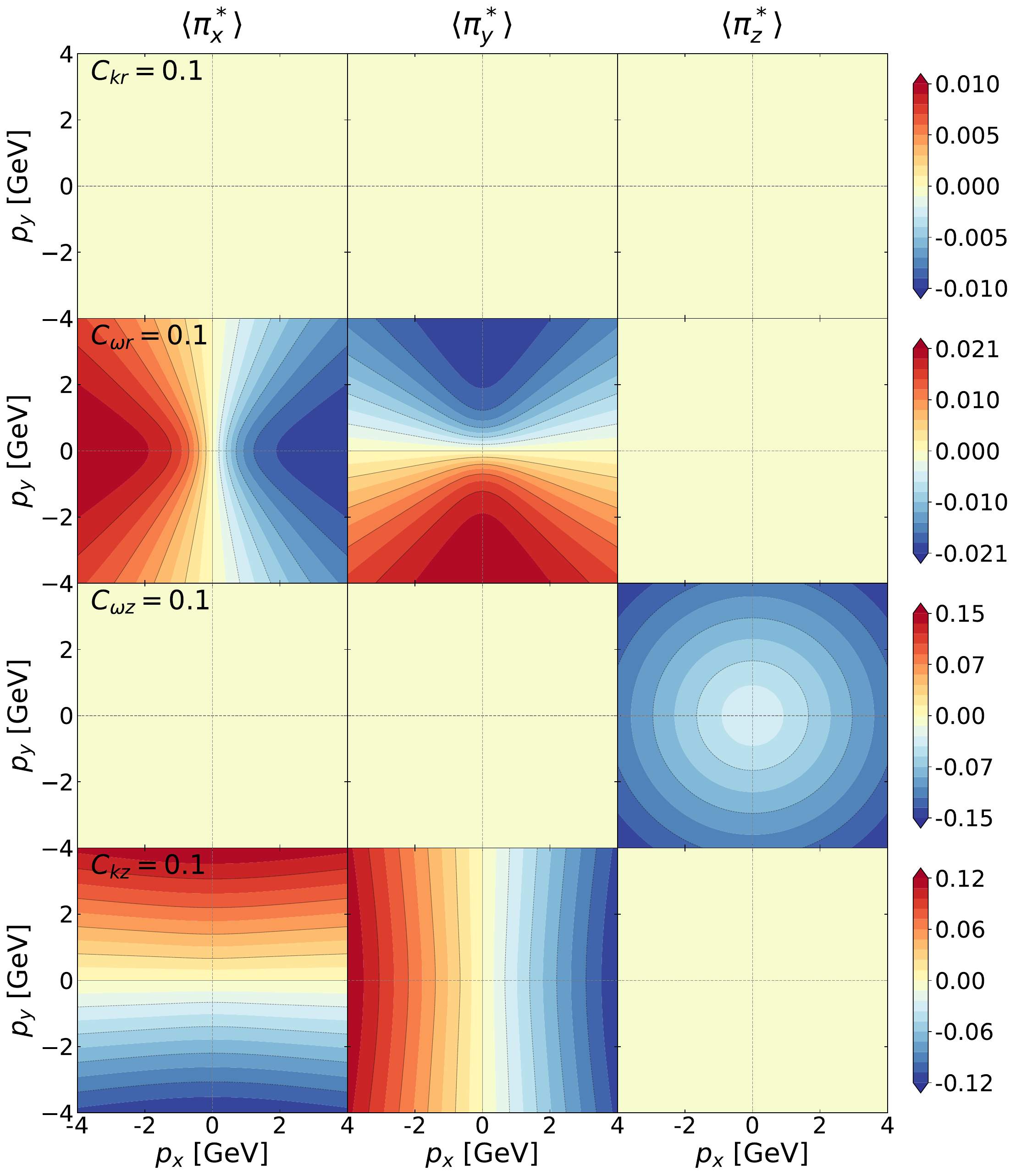}
    \caption{Momentum dependence of the mean spin polarization vector components $\langle \pi_x^* \rangle$, $\langle \pi_y^* \rangle$, and $\langle \pi_z^* \rangle$ in the transverse momentum plane $(p_x, p_y)$, evaluated for a heavy particle mass of $m = 1$~GeV at the freeze-out temperature $T = 0.15$~GeV. The columns correspond to the Cartesian components of the polarization vector in the particle rest frame. The four rows display the results for the four independent initial configurations: $C_{kr} = 0.1\rho(r)$ (first row), $C_{\omega r} = 0.1\rho(r)$ (second row), $C_{\omega z} = 0.1\rho(r)$ (third row), and $C_{kz} = 0.1\rho(r)$ (fourth row). Empty panels indicate exactly vanishing polarization components due to the underlying symmetry constraints.}
    \label{fig:Pol_Maps_1GeV}
\end{figure*}

\begin{figure*}[h!]
    \centering
    \includegraphics[width=0.95\textwidth]{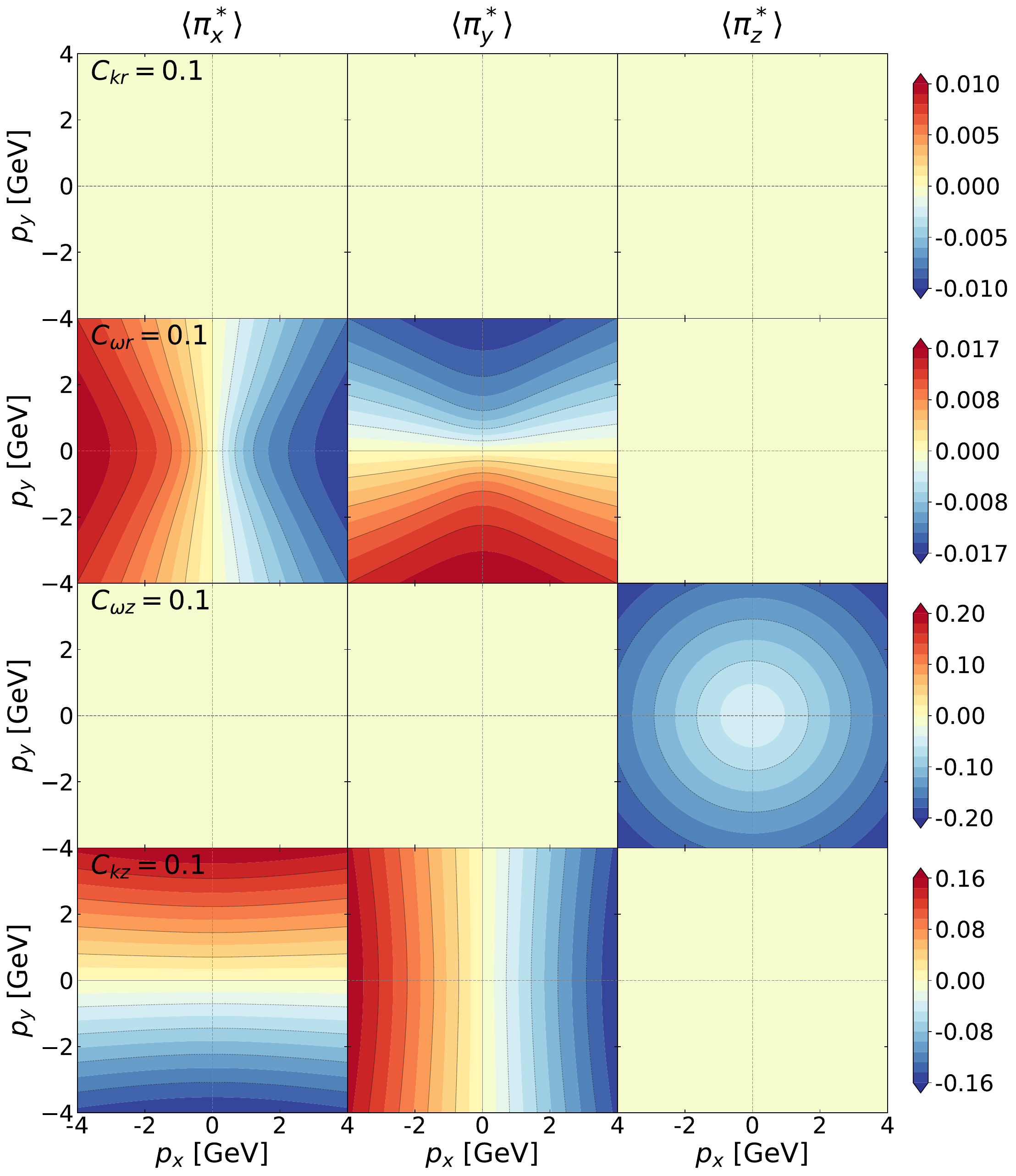}
    \caption{Momentum dependence of the mean spin polarization vector components $\langle \pi_x^* \rangle$, $\langle \pi_y^* \rangle$, and $\langle \pi_z^* \rangle$ in the transverse momentum plane $(p_x, p_y)$, evaluated for a lighter particle mass of $m = 0.3$~GeV at the freeze-out temperature $T = 0.15$~GeV. The layout of the panels and the initial conditions are identical to those in \FIG{fig:Pol_Maps_1GeV}.}
    \label{fig:Pol_Maps_0p3GeV}
\end{figure*}

Considering particles produced at mid-rapidity ($y=0$), the longitudinal momentum vanishes ($p_z = 0$). Consequently, the boost velocity $\bm{v}$ from the laboratory frame to the PRF is purely transverse. Its Cartesian components are numerically defined as $\bm{v} = (p_x/m_T, p_y/m_T, 0)$, and the corresponding Lorentz factor simplifies directly to the ratio $\gamma = m_T / m$. Utilizing these specific kinematic variables, the spatial components of the mean polarization vector in the PRF are derived through the standard Lorentz boost
\begin{equation}
    \langle \bm{\pi}^* \rangle = \langle \bm{\pi}_{\rm LAB} \rangle + \frac{\gamma^2}{\gamma+1} \bm{v} (\bm{v} \cdot \langle \bm{\pi}_{\rm LAB} \rangle) - \gamma \bm{v} \langle \pi^0_{\rm LAB} \rangle.
\end{equation}
Numerically, this vector equation is explicitly decomposed into its Cartesian components, where the scalar product $\bm{v} \cdot \langle \bm{\pi}_{\rm LAB} \rangle = v_x \langle \pi_x^{\rm LAB} \rangle + v_y \langle \pi_y^{\rm LAB} \rangle$ is evaluated using the contravariant spatial components of the laboratory-frame vector. These resulting PRF spatial components, $\langle \pi_x^* \rangle$, $\langle \pi_y^* \rangle$, and $\langle \pi_z^* \rangle$, directly constitute the two-dimensional transverse momentum polarization maps presented in \FIG{fig:Pol_Maps_1GeV} and \FIG{fig:Pol_Maps_0p3GeV}.

Since the boost to PRF is done in the transverse plane, it does not affect the components $F_3$ and $G_3$. Hence,  one may conclude that the longitudinal ($z$) component of the PL four-vector will be expressed by $F_3$ and $G_3$. Because these components depend on $C_{\omega z}$ and $C_{k \phi}$ coefficients only, we conclude that the longitudunal component of the PL four-vector can be different from zero only in the case {\bf (d)} . This is supported by our numerical results presented in \FIG{fig:Pol_Maps_1GeV} and \FIG{fig:Pol_Maps_0p3GeV}. Moreover, since $F_3$ and $G_3$ are independent of the angle $\phi_p$ we expect that longitudunal component of the PL four-vector will be cylindrically symmetric in the transverse plane, which is again confirmed by our numerical results shown in \FIG{fig:Pol_Maps_1GeV} and \FIG{fig:Pol_Maps_0p3GeV}.

\section{Summary}
\label{sec:summary}

We have numerically solved boost-invariant and cylindrically symmetric equations of perfect spin hydrodynamics. The energy--momentum and spin tensors have been taken in the form that describes a relativistic massive gas governed by Boltzmann statistics. We have found that radial expansion leads to a coupling between the azimuthal and longitudinal components of the electric and magnetic components of the spin polarization tensor. Although this feature was found earlier for the Gubser symmetry,  our treatment allows for a more general form of initial conditions and expansion geometry. Moreover, it yields regular solutions that are finite at the origin of the coordinate system. Defining the freeze-out hypersurface by the constant temperature condition, we have evaluated the Pauli--Lubański four-vector and found that for the assumed geometry the only nonzero total polarization may be induced by the longitudinal component of the magnetic part of the spin polarization tensor.

\medskip
\begin{acknowledgments}

We thank Radoslaw Ryblewski for useful discussions. This work was supported in part by the National Science Centre, Poland (NCN) Grant No.~2022/47/B/ST2/01372. 

\end{acknowledgments}

\begin{appendix}

\section{Discussion of the $1/r$ terms}

Two of the six equations that follow from the conservation of spin contain terms proportional to $1/r$ that do not vanish in the limit $\theta \to 0$, namely, Eq.~\eqref{eq:rphi} contains
$\frac{A C_{k\phi}}{2 r}$, and Eq.~\eqref{eq:uz}
contains~$\frac{A C_{\omega\phi}}{2 r}$.

Let us introduce the following notation for the two sectors of the spin tensor \eqref{eq:Sglw} that are responsible for those terms,
\begin{equation}
S^{\lambda, \mu \nu}_{\rm \Delta k} = \frac{A}{2}\big(\Delta^{\lambda \mu} k^\nu - \Delta^{\lambda \nu}k^\mu \big),
\end{equation}
\begin{equation}
S^{\lambda, \mu \nu}_{\rm tu} = \frac{A}{2}\big(t^{\lambda\mu} U^\nu - t^{\lambda \nu} U^\mu\big).
\end{equation}
Then it can be verified that
\begin{equation}\label{eq:a11}
R_\mu\Phi_\nu\,\partial_\lambda S^{\lambda,\mu\nu}_{\Delta k} =
\frac{A}{2r} \left(\partial_\phi C_{kr}-C_{k\phi}\right),
\end{equation}
\begin{equation}\label{eq:a12}
U_\mu Z_\nu\,\partial_\lambda S^{\lambda,\mu\nu}_{tu} = \frac{A}{2r}\left(C_{\omega\phi}-\partial_\phi C_{\omega r}\right).
\end{equation}
If the coefficients $C_{kr}$ and $C_{\omega r}$ are independent of $\phi$, as we have assumed, then we are left with precisely the terms that we mentioned above \footnote{The flipped sign in~\eqref{eq:rphi} comes from the fact that Eq.~\eqref{eq:rphi}, as written, was multiplied by $-1$.}.

The fact that the boost-invariant transversely homogeneous equations considered in previous studies~\cite{Drogosz:2024lkx, Drogosz:2026qbo} are not recovered by setting $\theta \to 0$ is natural and comes from the fact that a 
vector field that is constant and nonzero in the transverse plane is not a special case of a cylindrically symmetric vector field. The geometries are different. To recover the previous case, one would have to change the setting by introducing an explicit dependence of the coefficients $C_i$ on $\phi$ and relate the coefficients considered in the present work to coefficients in a basis that is Cartesian in the transverse directions,
\bea
\Ckv = C_{kx} X^\mu + C_{ky} Y^\mu + C_{kz} Z^\mu , \qquad
\Cov = C_{\omega x} X^\mu + C_{\omega y} Y^\mu + C_{\omega z} Z^\mu,
\eea
via
\begin{equation}
C_{kr}(\phi)=C_{kx}\cos\phi+C_{ky}\sin\phi,
\qquad
C_{k\phi}(\phi)=-C_{kx}\sin\phi+C_{ky}\cos\phi,
\end{equation}
\begin{equation}
C_{\omega r}(\phi)=C_{\omega x}\cos\phi+C_{\omega y}\sin\phi,
\qquad
C_{\omega\phi}(\phi)=-C_{\omega x}\sin\phi+C_{\omega y}\cos\phi.
\end{equation}
Then one would have
\begin{equation}
\p_\phi C_{kr}=C_{k\phi}, \qquad \p_\phi C_{\omega r}=C_{\omega \phi},
\end{equation}
and both $1/r$ coefficients would vanish,
\begin{equation}
\partial_\phi C_{kr}-C_{k\phi}=0,
\qquad
C_{\omega\phi}-\partial_\phi C_{\omega r}=0.
\end{equation}

\section{Freeze-out hypersurface}

The four-vector $\Sigma^\mu$ that defines 
the freeze-out hypersurface via Eq.~(\ref{Sigma}) possesses a simple decomposition in the considered basis,
\beq
\Sigma^\mu =
\left[\tau(\zeta)\cosh\theta-r(\zeta)\sinh\theta\right]U^\mu
+ \left[r(\zeta)\cosh\theta-\tau(\zeta)\sinh\theta\right]R^\mu .
\eeq
Analogously, the volume element from Eq.~(\ref{dSigma}) can be written as
\beq
\dd\Sigma^\mu =
\left[
\left(\frac{dr}{d\zeta}\cosh\theta
-\frac{d\tau}{d\zeta}\sinh\theta\right)U^\mu
+ \left(\frac{d\tau}{d\zeta}\cosh\theta
-\frac{dr}{d\zeta}\sinh\theta\right)R^\mu
\right]
r(\zeta)\tau(\zeta) \dd\zeta\,\dd\eta\,\dd\phi .
\eeq

\section{Contractions of four-momentum with the dual spin polarization tensor}
\label{app:PL}

After straightforward algebraic manipulations we obtain the explicit expressions for the vector ${\tilde \omega}_{\mu\nu} p^\nu$ appearing in \EQn{PDPLV},
\beq
{\tilde \omega}_{0\nu} p^\nu &=& p_T \cosh (\eta ) \LSB C_{\omega r} \cos (\phi-\phi_p) - C_{\omega \phi} \cosh \theta \sin (\phi -\phi_p)\RSB \nn \\
&& + C_{\omega z} \LSB m_T \cosh \theta \sinh (y)-p_T \sinh (\eta ) \sinh \theta \cos(\phi -\phi_p) \RSB \nn \\
&& + \sinh \theta \LSB C_{k \phi} m_T \sinh (y)+C_{kz} p_T \cosh (\eta ) \sin (\phi -\phi_p) \RSB \nn \\
&& - p_T \sinh (\eta ) \LSB C_{k \phi} \cosh \theta \cos (\phi-\phi_p) + C_{k r} \sin (\phi-\phi_p) \RSB,
\label{eq:w0}
\eeq

\beq
{\tilde \omega}_{1\nu} p^\nu &=& m_T \cosh (y-\eta ) \LSB C_{\omega \phi} \cosh \theta \sin (\phi )-C_{\omega r} \cos (\phi ) \RSB \nn \\
&& -\sinh \theta
\LSB C_{\omega \phi} p_T \sin (\phi_p)+C_{\omega z} m_T \cos (\phi ) \sinh (y-\eta ) \RSB \nn \\
&& +\cosh \theta \LSB C_{k z} p_T \sin (\phi_p)-C_{k \phi} m_T \cos (\phi ) \sinh (y-\eta) \RSB \nn \\
&& -m_T \sin (\phi ) \LSB C_{k r} \sinh (y-\eta )+C_{k z} \sinh \theta \cosh (y-\eta ) \RSB,
\label{eq:w1}
\eeq

\beq
{\tilde \omega}_{2\nu} p^\nu &=& \sinh \theta \LSB C_{\omega \phi} p_T \cos (\phi_p)-C_{\omega z} m_T \sin (\phi ) \sinh (y-\eta) \RSB \nn \\
&& -m_T \cosh (y-\eta ) \LSB C_{\omega \phi} \cosh \theta \cos (\phi )+C_{\omega r} \sin (\phi ) \RSB \nn \\
&& + m_T \cos (\phi ) \LSB C_{k r} \sinh (y-\eta )+C_{k z} \sinh \theta \cosh (y-\eta) \RSB \nn \\
&& -\cosh \theta \LSB C_{k \phi} m_T \sin (\phi ) \sinh (y-\eta )+C_{k z} p_T \cos (\phi_p) \RSB,
\label{eq:w2}
\eeq

\beq
{\tilde \omega}_{3\nu} p^\nu &=& p_T \sinh (\eta ) \LSB C_{\omega \phi} \cosh \theta \sin (\phi -\phi_p)-C_{\omega r} \cos (\phi
   -\phi_p) \RSB  \nn \\
&& -C_{\omega z} \LSB m_T \cosh \theta \cosh (y) - p_T \cosh (\eta ) \sinh \theta \cos (\phi -\phi_p) \RSB \nn \\
&& +p_T \cosh (\eta ) \LSB C_{k \phi} \cosh \theta \cos (\phi -\phi_p)+C_{k r} \sin (\phi
   -\phi_p) \RSB \nn \\
&& -\sinh \theta \LSB C_{k \phi} m_T \cosh (y)+C_{k z} p_T \sinh (\eta ) \sin(\phi -\phi_p) \RSB.
\label{eq:w3}
\eeq

In the case $\theta = \phi = 0$, \EQSM{eq:w0}{eq:w3} reduce to Eq.~(61) in \CITn{Florkowski:2019qdp}, if we make identifications $R^\mu=X^\mu$ and $\Phi^\mu=Y^\mu$.

\section{Boost to PRF}
\label{app:FL}

\subsection{Case {\bf (a)}}
\label{appBa}

If only the $C_{k r}$ component is different from zero, one can check that $F^*_\mu = 0$
and $G^*_\mu = 0$, which is confirmed by the numerical results presented in \FIG{fig:Pol_Maps_1GeV} and \FIG{fig:Pol_Maps_0p3GeV} (the top raws show that all the components of the Pauli--Lubański four-vector vanish).

\subsection{Case {\bf (b)}}
\label{appBb}

In this case, the only nonzero component of the spin polarization tensor is $C_{\omega r}$. The boost to the particle rest frame gives
\beq
F^*_\mu &=& 4 \pi C_{\omega r} I_1 (\pgoth) (K_0 (\mgoth) m_T + K_1 (\mgoth) T \sech\theta)  \f{m}{m_T} \,\, (0, -\cos \phi_p, -\sin \phi_p, 0 ) 
\eeq
and
\beq
G^*_\mu &=& 4 \pi C_{\omega r} K_1 (\mgoth) (I_2 (\pgoth) p_T + I_1 (\pgoth) T \csch\theta) \f{m}{p_T}
(0, -\cos \phi_p, -\sin \phi_p, 0 )  .
\eeq
For conciseness, the arguments of the Bessel functions use the notation introduced in Eq.~(\ref{eq:gothic}).

\subsection{Case {\bf (c)}}
\label{appBd}

In this case the only nonzero components of the spin polarization tensor are $C_{\omega z}$ and $C_{k \phi}$. The boost to PRF gives
\beq
F^*_0 &=& F^*_1 = F^*_2 = 0, \nn \\
F^*_3 &=& 2 \pi  \left[ \sinh (\theta ) \left( C_{\omega z} I_1 (\pgoth) \kappa \, p_T - 2 C_{k \phi} I_0 (\pgoth) K_1 (\mgoth) m_T \right) \right.
\nn \\
&& \left. +\cosh (\theta ) \left( C_{k \phi} I_1 (\pgoth) \kappa \,
    p_T-2 C_{\omega z} I_0 (\pgoth) K_1 (\mgoth) m_T \right) \right]
\eeq
and
\beq
G^*_0 &=& G^*_1 = G^*_2 = 0, \nn \\
G^*_3 &=&  4 \pi \sinh(\theta ) \left[ K_1 (\mgoth) (C_{k \phi} \coth (\theta ) + C_{\omega z}) (I_1 (\pgoth) T \text{csch}(\theta )+ I_2 (\pgoth) p_T) \right. \nn \\
&& \left. -I_1 (\pgoth) K_0 (\mgoth) m_T
   \left( C_{k \phi}+C_{\omega z} \coth (\theta ) \right) \right].
\eeq

\subsection{Case {\bf (d)}}
\label{appBc}

In this case, the only nonzero components are $C_{k z}$ and $C_{\omega \phi}$.  A straightforward calculation gives
\beq
F^*_0 &=& 0, \nn \\
F^*_1 &=& 
4 \pi \sin (\phi_p) \LSB I_0 (\pgoth) K_1 (\mgoth) p_T \LSB C_{k z} \cosh (\theta )-C_{\omega \phi} \sinh (\theta ) \RSB
\right. \nn \\
&& \left.
+I_1 (\pgoth) \text{sech}(\theta ) \LSB C_{\omega \phi}
\cosh (\theta) - C_{k z} \sinh (\theta ) \RSB  (K_0 (\mgoth) m_T \cosh (\theta )+K_1 (\mgoth) T) \RSB
, \nn \\
F^*_2 &=& 
4 \pi \cos (\phi_p) (I_0 (\pgoth) K_1 (\mgoth) p_T (C_{\omega \phi} \sinh (\theta )-C_{k z} \cosh (\theta )) \nn \\
&& -I_1 (\pgoth) (C_{\omega \phi}-C_{k z} \tanh (\theta
   )) (K_0 (\mgoth) m_T \cosh (\theta )+K_1 (\mgoth) T))
   , \nn \\
F^*_3 &=& 0,
\eeq
and
\beq
G^*_0 &=& 0, \nn \\
G^*_1 &=& \f{4 \pi}{p_T} \sin (\phi_p) [-p_T \sinh (\theta ) (C_{k z} I_2 (\pgoth) K_1 (\mgoth) m_T+C_{\omega \phi} I_1 (\pgoth) K_0(\mgoth) p_T)
\nn \\
&& 
+p_T \cosh (\theta) (C_{k z} I_1 (\pgoth) K_0 (\mgoth) p_T+C_{\omega \phi} I_2 (\pgoth) K_1 (\mgoth) m_T) \nn \\
&&
+I_1 (\pgoth) K_1 (\mgoth) m_T T (C_{\omega \phi} \coth (\theta
   )-C_{k z})]
   , \nn \\
G^*_2 &=& \f{4 \pi}{p_T} \cos (\phi_p) [ p_T \sinh (\theta ) (C_{k z} I_2 (\pgoth) K_1 (\mgoth) m_T+C_{\omega \phi} I_1 (\pgoth) K_0 (\mgoth) p_T)\nn \\
&& -p_T \cosh (\theta )
   (C_{k z} I_1 (\pgoth) K_0 (\mgoth) p_T+C_{\omega \phi} I_2 (\pgoth) K_1 (\mgoth) m_T) \nn\\
&&
+I_1 (\pgoth) K_1 (\mgoth) m_T T (C_{k z}-C_{\omega \phi} \coth (\theta)) ]
   , \nn \\
G^*_3 &=& 0.
\eeq

\end{appendix}

\bibliographystyle{apsrev4-1}
%

\end{document}